\documentclass[11pt,oneside,letterpaper]{article}

\usepackage{etex} 
\usepackage{amssymb,amsfonts,amsmath,amstext}
\usepackage{natbib}
\usepackage{bibentry}    
\usepackage{setspace}
\usepackage{multirow}      
\usepackage{booktabs}      
\usepackage{bbm}           
\usepackage{enumitem}

\usepackage{graphicx} 
\usepackage[update,prepend]{epstopdf}
\graphicspath{{../v_common/graphs/}}
\usepackage{hyperref}
\hypersetup{
  unicode=false,          
  pdftoolbar=true,        
  pdfmenubar=true,        
  pdffitwindow=false,     
  pdfstartview={FitH},    
  pdftitle={Discrete Games in Endogenous Networks}, 
  pdfauthor={Anton Badev},      
  pdfsubject={Subject},   
  pdfcreator={Creator},   
  pdfproducer={Producer}, 
  pdfnewwindow=true,      
  colorlinks=true,        
  linkcolor=blue,         
  citecolor=black,        
  filecolor=brown,        
  urlcolor=black          
}

\usepackage{tikz}
\usetikzlibrary{arrows,shapes,positioning,backgrounds,calc,trees}
\pgfdeclarelayer{background}
\pgfsetlayers{background,main}

\usepackage[amsmath,hyperref,framed,standard]{ntheorem} 
\usepackage{framed}     
\usepackage{pstricks}   
\theoremclass{Theorem}
\theoremstyle{plain}
\theoremheaderfont{\normalfont\bfseries}
\theorembodyfont{\itshape}

\renewtheorem{theorem}[Theorem]{Theorem}
\renewtheorem{definition}[Definition]{Definition}
\renewtheorem{lemma}[Lemma]{Lemma}
\renewtheorem{proposition}[Proposition]{Proposition}
\renewtheorem{corollary}[Corollary]{Corollary}

\theoremclass{Assumption}
\newtheorem{assumption}{Assumption}
\newtheorem{claim}{Claim}

\newcommand{\qed}{\hfill\rule{0.4em}{0.4em}}

\newcommand{\comments}[1]{}

\newcommand{\bit}[1]{{\noindent\emph{{#1}}}}

\newcommand{\bsc}[1]{\vspace{0.2cm}{\noindent\textsc{{#1}}}}

\newcommand{\Pfn}{\mathcal{P}}               
\newcommand{\Gn}{\mathbf{S}}                 
\newcommand{\SSn}{\mathbf{S}_n}                 
\newcommand{\Sn}{\mathbf{S}}                 
\newcommand{\Si}{\mathbf{S}_{(i)}}           
\newcommand{\setN}{\mathbf{N}}           

\newcommand{\dSSn}{\Delta(\mathbf{S}_n)}                 

\newcommand{\Xn}{\mathbf{X}}        
\newcommand{\KCS}{S^*_{k-CS}}       
\newcommand{\KNS}{S^*_{k-NS}}       
\newcommand{\TWONS}{S^*_{2-NS}}     
\newcommand{\NNS}{S^*_{n-NS}}       
\newcommand{\Smode}{S^{mode}}       

\newcommand{\NS}{\mathbf{N}}        
\newcommand{\Ns}{\NS}        
\newcommand{\MK}{\mathbf{M}}        
\newcommand{\KK}{\mathbf{K}}        

\newcommand{\E}{\mathbb{E}}
\newcommand{\R}{\mathbb{R}}

\newcommand{\ind}{\mathbbm{1}}

\newcommand{\e}{\mathrm{e}}

\newcounter{countnoteitem}

\newcommand{\fignotetitle}[1]{\small \textit{#1}}
\newcommand{\fignotetext}[1]{\small {#1}}
\newcommand{\axislabel}[1]{\footnotesize \textit{#1}}

\definecolor{UniBlue}{RGB}{83,121,170}
\definecolor{lavander}{cmyk}{0,0.48,0,0}
\definecolor{violet}{cmyk}{0.79,0.88,0,0}
\definecolor{burntorange}{cmyk}{0,0.52,1,0}
\definecolor{rosybrown}{rgb}{0.74,0.56,0.56}

\def\nodeborder{orange}
\def\nodetext{black}
\def\nodefont{\bfseries\tiny}
\def\linkfont{\color{rosybrown} \scriptsize}
\def\node_index_color{\color{black}}
\def\inf_node_background_color{red!70}

\tikzstyle{node}=%
[draw,circle,inner sep=2.0pt,minimum width=5pt,
thick,\nodeborder,text=\nodetext,font=\nodefont]

\tikzstyle{inf node}=%
[draw,star,star points=5,inner sep=0.8pt,minimum width=17pt,
thick,\nodeborder,text=\nodetext,font=\nodefont,
bottom color=\inf_node_background_color, top color=white]

\tikzstyle{inf node a}=%
[draw,diamond,inner sep=0.8pt,minimum width=17pt,
thick,\nodeborder,text=\nodetext,font=\nodefont]

\tikzstyle{inf node b}=%
[draw,star,star points=5,inner sep=0.8pt,minimum width=17pt,
thick,\nodeborder,text=\nodetext,font=\nodefont]

\tikzstyle{legend_node}=%
[circle, draw, \nodeborder, rounded corners,thick,
bottom color=white, top color=white,
text=\nodetext,font=\tiny\bfseries,inner sep=0.0pt,text width=0.8cm,minimum width= 1.1cm,align=center]

\tikzstyle{legend_inf_node}=%
[star, star points=5, draw, \nodeborder, 
bottom color=\inf_node_background_color, top color= white,
text= \nodetext,font=\tiny\bfseries,inner sep=0.0pt,text width=0.6cm,minimum width= 0.9cm,align=center]

\tikzset{
    >=stealth',
    punkt/.style={
           rectangle,
           rounded corners,
           draw=black, very thick,
           text width=6.5em,
           minimum height=2em,
           text centered},
    pil/.style={
           ->,
           thick,
           shorten <=2pt,
           shorten >=2pt,}
}

\newcommand{\convexpath}[2]{
[
    create hullnodes/.code={
        \global\edef\namelist{#1}
        \foreach [count=\counter] \nodename in \namelist {
            \global\edef\numberofnodes{\counter}
            \node at (\nodename) [draw=none,name=hullnode\counter] {};
        }
        \node at (hullnode\numberofnodes) [name=hullnode0,draw=none] {};
        \pgfmathtruncatemacro\lastnumber{\numberofnodes+1}
        \node at (hullnode1) [name=hullnode\lastnumber,draw=none] {};
    },
    create hullnodes
]
($(hullnode1)!#2!-90:(hullnode0)$)
\foreach [
    evaluate=\currentnode as \previousnode using \currentnode-1,
    evaluate=\currentnode as \nextnode using \currentnode+1
    ] \currentnode in {1,...,\numberofnodes} {
-- ($(hullnode\currentnode)!#2!-90:(hullnode\previousnode)$)
  let \p1 = ($(hullnode\currentnode)!#2!-90:(hullnode\previousnode) - (hullnode\currentnode)$),
    \n1 = {atan2(\y1,\x1)},
    \p2 = ($(hullnode\currentnode)!#2!90:(hullnode\nextnode) - (hullnode\currentnode)$),
    \n2 = {atan2(\y2,\x2)},
    \n{delta} = {-Mod(\n1-\n2,360)}
  in
    {arc [start angle=\n1, delta angle=\n{delta}, radius=#2]}
}
-- cycle
}

\usepackage[margin=1.2in]{geometry}
\begin{document}

\title{Discrete Games in Endogenous Networks: Equilibria and Policy%
\footnote{
Based on my Ph.D. dissertation (\citealp{badev2013discrete}) at the University of Pennsylvania under the guidance of Kenneth Wolpin, George Mailath and Petra Todd. I have benefited from discussions with Steven Durlauf, Hanming Fang,  James Heckman, Matt Jackson, Ali Jadbabaie, Michael Kearns, Angelo Mele, Antonio Merlo, \`{A}ureo de Paula and seminar audiences at Bocconi, Chicago (2012 SSSI), 2012 QME (Duke), GWU, 2015 NSF-ITN (Harvard), Mannheim, Minnesota, 2012 NASM (Northwestern), NYU, Penn, Pitt/CMU, St. Louis (Econ and Olin), 2015 ESWC (Montreal), 2016 SITE (Stanford), 2014 SED (Toronto), 2014 EMES (Toulouse), Tilburg, Texas Tech, University of Hawaii, 2015 IAAE, 2016 AMES (Kyoto), and Yale (SOM) for useful comments.
I gratefully acknowledge financial support from the TRIO (PARC/Boettner/NICHD) Pilot Project Competition. This work used the XSEDE, which is supported by National Science Foundation grant number OCI-1053575. All errors are mine.%
}
\\
}
\author{Anton Badev%
\thanks{The views expressed herein are those of the authors and not necessarily those of the Board of Governors of the Federal Reserve System.} \\
Board of Governors and Tepper School of Business, CMU
}
\date{ 
5/8/2017 
}
\maketitle

\renewcommand{\abstractname}{}
\vspace{-1.1cm}
\begin{abstract}
  \noindent \textbf{Abstract.}
  I postulate that social norms and individuals' behaviors are shaped by a common process capable of generating a multitude of outcomes. In games of friendship links and behaviors, I propose $k$-player Nash stability---a family of equilibria, indexed by a measure of robustness given by the number of permitted link changes, which is (ordinally and cardinally) ranked in a probabilistic sense. Application of the proposed framework to adolescents' tobacco smoking and friendship decisions suggests that: (a.) friendship networks respond to increases of tobacco prices and this response amplifies the intended policy effect on smoking, (b.) racially desegregating high-schools, via stimulating the social interactions of students with different intrinsic propensity to smoke, decreases the overall smoking prevalence, (c.) adolescents are averse to sharing friends so that there is a rivalry for friendships, (d.) when data on individuals' friendship network is not available, the importance of price centered policy tools is underestimated.

\end{abstract}
\noindent \textbf{JEL Codes:} D85, C73, L19 \\
\noindent \textit{Keywords:} Social Networks, Adolescent Smoking, Multiplicity, Discrete Games.%

\newpage
\setstretch{1.2}

\newpage
\renewcommand{\thepage}{\arabic{page}}

\section{Introduction}
Arguably human beings are both products and creators of their social environments. Humans engage in a complex system of social relations, also known as social interactions, which determine their social environment, influence their socio-economic choices, and affect the market outcomes. It has been largely overlooked, however, that the converse is also true: humans' behaviors also shape their social environment, and this paper explores the two-way interplay between individuals' behaviors and individuals' social environment defined by their friendship network. 

I postulate that social norms and behaviors are shaped \emph{jointly} by a common process capable of generating a \emph{wide} range of outcomes. The complexity of this process is naturally captured by a game of link and node statuses, and this paper contributes the understanding of the (non-linear) interplay between individuals' decisions to form friendships and engage in risky behaviors in three dimensions: (i) by proposing a general family of equilibria--outcomes which are likely to arise and persists---in settings where individuals jointly choose actions and linking strategies, (ii) by obtaining both ordinal and cardinal probabilistic ranking of these equilibria which ranking has broader implications including for both the estimation of and simulations from the proposed model, and (iii) by presenting empirical evidence that: (a.) friendship networks respond to policies narrowly targeting tobacco smoking such as tobacco price increase, and this response amplifies the intended policy effects on smoking, (b.) racially desegregating high-schools, via stimulating the social interactions of students with widely different intrinsic propensity to smoke, decreases the overall smoking prevalence, (c.) adolescents are averse to sharing common friends so that there is a rivalry for friendships, (d.) the estimates of the effect of price on tobacco smoking is understated in both cases when data on individuals' friendship network is not available or when the model does not account for the response of the friendship network to changes in tobacco prices.

Friends' influences have often been pointed to as a major driver of human behaviors and have been associated with the potential to create a domino effect.\footnote{See \cite{Leibenstein1950} for an early discussion.} The main premise of this paper is that it it possible a converse mechanism to be in place where a change of individual A decisions re-shapes who her peers are as opposed to pressure her peers to follow A's decisions, e.g. an individual who ceases smoking (after an increase of tobacco prices) may reconsider her friendship network! To analyze the potential of this two-way effect then we need a model in which individuals decide on both their friendships and behaviors (decisions to smoke). Importantly, this model ought to acknowledge the possible externalities between both smokers and their peers, and also between peers themselves, e.g. peers may be rivals for friends. Relatedly and in addition, the model will need a coherent notion of equilibrium adequately describing agents behavior (in these settings) which, ideally, is suitable for inference. In short, a suitable model ought to handle explicitly (vs assume away) the possibility for multiple equilibria, inherited from the presence of externalities, and the complexity of the equilibrium play, inherited from the complexity of the decision environment.



In a friendship network, if agents make choices in their best interest, which patterns of links and behaviors are likely to prevail? Given an agent's incentives, her observed links and behaviors are likely to compare favorably against her alternatives, i.e. are likely to be robust against a set of feasible deviations. Considerations related to the complexity of individuals' decision problem could lend support to restricting the set of permissible deviations to local ones.\footnote{There are $2^{n-1}$ possible link deviations and only $n-1$ possible one-link-at-a-time link deviations. \cite{Jackson_Wolinsky_1996} propose a notion of stability where the presence of a link implies that none of the parties involved has unilateral incentive to sever this link. Though, severing multiple links are not contemplated.} Reasoning about the \emph{diameter} of these local deviations, however, has not fallen in the focus of extant research and motivates us to introduce a class of (non-cooperative) equilibria indexed by the diameter of the permissible local deviations. For a population of size $n$, \emph{$k$-player Nash stability} ($k$-PS) as a configuration of links and node statuses where any subset of $k$ players is in a Nash equilibrium of the induced game between these $k$ players when only the friendships between the $k$ players are decided together with their action statuses. Consequently, in a $k$-player Nash stable network state, no player has an incentive to alter simultaneously $k-1$ of his friendships and his action. For $k<n$, $k$-PS is less demanding on the players in comparison with a Nash play (i.e., when $k=n$) and is, therefore, a more tenable assumption in large population games. When $k$ increases from 2 to $n-1$, the $k$-PS family gradually fills in the gap between the local stability argument of \cite{Jackson_Wolinsky_1996} and the Nash play discussed in \cite{Bala_Goyal_2000}.

\comments{However, $k$-player Nash stability suffers well known deficiencies, on which \cite*{kandori_mailath_rob_93} remark:
\begin{quote}
While the Nash equilibrium concept has been used extensively in many diverse contexts, game theory has been unsuccessful in explaining \emph{how} players know that a Nash equilibrium will be played. Moreover, the traditional theory is silent on how players know \emph{which} Nash equilibrium is to be played if a game has multiple equally plausible Nash equilibria.
\end{quote}
}

In parallel to $k$-PS, I introduce the \emph{$k$-player dynamic} ($k$-PD), a family of myopic dynamic processes which not only offers a ``mass action'' interpretation of $k$-PS (as opposed to $k$-PS arising from a complex reasoning process, \citealp{Nash1950}) but also delivers, cardinal and ordinal, probabilistic ranking of the entire equilibrium family.\footnote{See the seminal papers of \cite{foster_young_90}, \cite{kandori_mailath_rob_93,Blume_1993,JacksonWatts2002}. Some of the ideas I exploit are encountered in \citet[Chapter VII]{Cournot_1838}.} In a $k$-PD, every period an individual meets $k-1$ potential friends and decides whether or not to befriend each of them as well as whether to revise her action choice. The $k$-PD family induces a stationary distribution over the entire set of possible outcomes which \emph{embeds} the family of $k$-PS in an intuitive way (each $k$-PS is $k$-neighborhood local mode of the stationary distribution) and which, because of its invariance to $k$, \emph{ranks} probabilistically each equilibrium within the family, even for different $k$-s. In addition to the cardinal ranking, the analysis of the $k$-PD independently delivers, as a by-product, a re-affirmative ordinal ranking of these equilibria. The larger $k$ is, the faster $k$-PDs approach the stationary distribution, i.e. the more likely is the stationary distribution to represent those $k$-PDs, and the more probable the rest points of these $k$-PDs (of course $k$-PS states) are. 

The convergence properties of $k$-PD have immediate implications for the implementation of the proposed model. The model's likelihood is given by the (unique) stationary distribution of the $k$-PD family. This distribution pertains to the Exponential Random Graph Models\footnote{See \cite{FrankStrauss1986markov,WassermanPattison1996}}, for which both direct estimation and simulating from the model with known parameters are computationally infeasible.\footnote{An evaluation of the likelihood requires the calculation of a summation with $2^{n^2}$ terms, e.g. for $n=10$, $2^{100}$ terms. In addition, MCMC algorithms offer a poor approximation (\citealp{BhamidiBrestlerSly_2011}).} 
For these models the double Metropolis-Hastings sampler offers a Bayesian estimation strategy, which nevertheless rely on simulations from the stationary distribution via Markov chains.\footnote{See \cite{Liang2010double} and \cite{murray2012mcmc}. \cite{Mele2017} studies asymptotic approximations of this sampler and, in his numerical experiments, points to numerical benefits of occasionally using large update steps. The analysis of $k$-PD here offers a rigorous basis for this observation with a model based on economic behavior.} Importantly, in my settings, these Markov chains can be readily simulated via the $k$-PD family, whose varying convergence rates offer a way to break the poor convergence properties associated with local Markov chains.\footnote{In a local Markov chain each update is of size $o(n)$. For more see \cite{BhamidiBrestlerSly_2011}.} The particular novelty is that $k$-PD justifies sampling from the stationary distribution with \emph{varying} $k$ on the support $\{2, \ldots, n-1\}$, speeding the convergence of the proposed algorithm.

The model is estimated with data on smoking behavior, friendship networks, and home environment (parental education background and parental smoking behavior) from the National Longitudinal Study of Adolescent Health.\footnote{Details about the Add Health data, including the sample construction, are in the appendix.} 
This is a longitudinal study of a nationally representative sample of adolescents in the United States, who were in grades $7$--$12$ during the $1994$--$95$ school year. The estimation results confirm the leading role of the home environment in shaping adolescents' risky behaviors. Among the drivers of individuals decision a major factor is their home environment: the presence of a smoker in the household increases (on average) the likelihood of an adolescent smoking by $12.4$ ppt. In addition, if the student's mother has completed high school or college, this likelihood falls (on average) by $4.8$.

The estimation exercise also reveals an intricate non-linearities pertaining to the returns and rivalries from friendships. 
In particular, adolescents prefer exclusive friendships as opposed to sharing friends which is consistent with rivalry for time spent with common friends. Importantly, this rivalry can be uncovered only after the model accounts for the nonlinear returns to friendships, i.e. each additional friendship may bring increasing/decreasing marginal utility.\footnote{This observation is consistent with the fact that individuals with higher degrees participate in fewer clusters. I thank Mat Jackson for pointing to me this intuition.} In addition, estimates of the price coefficient on smoking which do not rely on data from the friendship network point to the presence of intricate (non-linear) relationships between smoking, peer norms and exogenous determinants of smoking such as tobacco prices. In particular, in these cases the price effect on smoking is understated (in fact insignificant).\footnote{Standard arguments signing the omitted variable bias cannot be invoked because the correlation between the (omitted) peer norm and price is of ambiguous sign. 
}

I use the estimated model to perform a number of counterfactual experiments. First, I ask whether the friendship network responds to changes in tobacco prices, e.g. an individual may stop smoking \emph{and} sever friendships with smokers, and whether this response has policy relevant (quantitative) implications. When comparing how individuals respond in fixed versus endogenous friendship network environments there are two effects to consider. The direct effect of changing tobacco prices is the first order response and, intuitively, will be larger whenever individuals are free to change their friendships, i.e. more individuals are likely to immediately respond to changes in tobacco prices provided they are not confined to their (smoking) friends. The indirect (ripple) effect of changing tobacco prices is the effect on smoking which is due, in part, to the fact that one's friends have stopped smoking. Contrary to before, a fixed network propels the indirect effect, e.g. an individual who changes her smoking status is bound to exert pressure to her friends (most likely smokers) and, thus, likely to alter her friends' decisions to smoke. It is, then, an empirical question how these two opposing effects balance out. Simulations with the estimated model suggest that following an increase in tobacco prices, the direct effect dominates, i.e. the response of the friendship network amplifies the intended policy effects in the neighborhood of $10\%$.

Next, I analyze how changes in the racial composition of schools affect adolescent smoking. When students from different racial backgrounds study in the same school, they interact and are likely to become friends. Being from different racial backgrounds student have different intrinsic propensity to smoke and the question is what is the equilibrium behavior in these mixed-race friendships--those who do not smoke start smoking or those who smoke stop smoking. Simulations from the model suggest that redistributing students from racially segregated schools into racially balanced schools decreases the overall smoking prevalence by slightly above $10\%$. 

Finally, for a school from the sample with a particularly high smoking prevalence ($45\%$), I examine the possibility of an intervention targeting only part of this school's population which is capable of changing their smoking decisions\footnote{The policy may consists of providing direct incentives or information about the health risks associated with smoking tobacco. Then, it may be too expensive to treat the entire school and, instead, the policy maker may engage only a small part of the school with the purpose to alter the social norms.}. The (empirical) question of interest is when treated individuals return in the schools, will their friends follow their example, i.e. extending the effect of the proposed policy beyond the set of treated individuals and thus creating a domino effect, or will their, previous to the treatment, friends un-friend them? In essence this is a question about the magnitude of the spillover effects and this paper contribution is to account for the possibility of the friendship network to adjust to the proposed treatment. Indeed, simulations with the model reveal that this spillover effect could be in the neighborhood of $5$ folds.\footnote{Note that in the absence of social interactions, this effect will be absent as well.}

\subsection*{Related literature}

The work of \cite{Nakajima2007} and \cite{Mele2017} have provided direct inspiration for this research. \cite{Nakajima2007} empirically studies peer effects abstracting from the friendship formation and \cite{Mele2017} obtains large network asymptotics of the model when restricted to the link formation only. A handful of theoretical papers consider (broadly related) adaptive link dynamics or model both network formation along with other choices potentially affected by the network (See \citealp{JacksonWatts2002,goyal_vega-redondo_05,Cabrales2011,Hiller2012,koenig_tessone_zenou_12,Baetz2014}). The theoretical frameworks available, however, are meant to provide focused insights into isolated features of networks and deliver sharp predictions, but are not easily adapted for the purposes of estimation. In addition, a typical approach is to focus the analysis on a particular equilibrium as opposed to discussing all equilibria. Multiplicity of equilibria reflects the possibility of network and behavioral externalities which is an indispensable feature of our settings. 



Contemporary papers to propose econometric treatment of models of networks and actions are those of \cite{goldsmith2013social} and \cite{hsieh2014social}. The focus of their work, however, is not policy analysis, which permits them to avoid the equilibrium microfoundations of a strategic model and an explicit treatment of the possibility for multiplicity.\footnote{Multiplicity in our settings makes the model intractable and, more importantly, deteriorates the statistical power of model's predictions.} Related work by \cite{hsieha2016network} propose a two-stage estimation procedure, with an application to R and D, which relies on conditional independence of links delivered by abstracting from link externalities. \cite{auerbach2016identification} obtains identification results within large network asymptotics which also rely on conditional independence of links.\footnote{While the assumptions leading to conditional independence present conveniences at the implementation stage, these limit the scope for studying peer effects, in addition ruling out nuanced motives such as friendship rivalries or non-linear returns to friendships. As argued above and demonstrated by the empirical analysis, these are of critical for studying the price effects on smoking.} Finally, there are recent contributions to the econometrics literature which focus on link formation, though are not easily extendable to include action choice as well.\footnote{See \cite{Sheng2014structural}, \cite{ArunJackson2016}, \cite{Leung2014random}, \cite{dePaulaRichardsTamer2014identification}, \cite{Graham2014econometric}, \cite{Menzel2015strategic} and the reviews in \cite{Arun2015econometrics,dePaula2016Econometrics,Bramoulle2016Oxford}.}

The empirical analysis of friendship networks and smoking behaviors lends support to a host of results which are related to the large body of empirical work on social interactions and teen risky behaviors. Of this literature, the closest to this work is \cite{Nakajima2007}. In terms of estimates, this paper makes the first step in explaining of how unavailability of data on the social network and not accounting for the response of these social network could contribute the generally low estimates for the price elasticity of smoking.\footnote{See \cite{ChaloupkaHenry1997} and \cite[Surgeon General's Report]{CDCSurgeonGeneral2000}.} In addition, existing empirical studies are confined to analysis with no data on friendship network or take the friendship network as given and focus on the identification of peer effects. In this sense, the premise of this research is that friends cannot be assigned so that one cannot ask by how much an additional friend who is a smoker increases one's propensity to smoke. 

The rest of the paper is organized as follows. Section 2 develops the theoretical analysis and obtains properties which are critical for the empirical implementation. Section 3 presents the data and estimation of the model, and section 4 discusses the implications from the countercfactual experiments. All formal proofs are in the appendix.

\section{A Model of Friendships and Behaviours} \label{section:theory}
The model is developed in two stages. First, agents' strategic behavior is analyzed in static settings and then a family of myopic dynamic processes is used to approximate the predictions of the static model in a inferentially suitable way. I start with a simple model prototype which, later on in our empirical application, is adapted to a more elaborate specification. 

\subsection{Players and preferences}


A finite population $\setN=\left\{1,2,...,n\right\}$ of agents decide on a binary action $a_{i}\in \{ 0,1\}$ and (asymmetric) relationships between each other $g_{ij}\in \{ 0,1\}$ for $i,j\in \setN$.\footnote{Our analysis trivially extends to continuous actions.} In our empirical application, $\setN$ is the collection of all student cohorts in a given high school at a given time period, where $a_{i}=1$ if student $i$ smokes and $g_{ij}=1$ if $i$ nominates $j$ as a friend.\footnote{Alternatively, the population of a geographically isolated area. Any closed collection of individuals who draw friends from within themselves will fit the assumptions of the model. Alternatively, this assumption (if the data do not span the complete network, for example) can be relaxed if one conditions on the existing friendships with outsiders of the pool. Peers who are not in the data are not part of the model, in the sense that the links to them can be thought of as part of the fixed attributes $X_i$. Then the model will explain the formation of new friendships conditioning on the ones with outsiders.} Agent $i$ selects her link nominations and action status $S_{(i)} = (a_i,\{g_{ij}\}_{j\neq i})$ from her choice set $\Si$ to maximize her payoff, which depends both on her exogenous characteristics $X_i$, e.g. age, gender, etc, and on her endogenous characteristics, e.g. network position, decision of her network neighbors, and etc. These endogenous characteristics capture externalities, in various contexts referred to as peer effects or spillovers, and are of direct interest to our study. 

The payoff of individual $i$, $u_i: \Sn \times \Xn\longrightarrow \R$, orders the outcomes in $\Sn$ given the attributes of the population $X=(X_1,...,X_n)\in \Xn$ in the following way:
\begin{eqnarray} \label{payoff}
u_i (S,X) & = & a_i v(X_i) + a_i h \sum_j a_j + a_i \phi \sum_j g_{ij}g_{ji} a_j\\ \notag
&   & + \sum_j g_{ij} w(X_i,X_j) + m \sum_j g_{ij}g_{ji} 
+ q \sum_{j,k} g_{ij}g_{jk}g_{ki}
\end{eqnarray}
Here $v(.)$ and $w(.,.)$ are functions of agents' (exogenous) characteristics. The incentives encoded in ($\ref{payoff}$) are easier to understand, if we consider the incremental payoffs from changing just one dimension $i$'s decision, i.e., the incremental payoff of changing $i$'s action status or a single link $g_{ij}$.

Consider $\Delta_{a_i}u_i(S,X) = u_i(a_{i}=1,S_{-i},X) - u_i(a_{i}=0,S_{-i},X)$ given as:
\begin{eqnarray} \label{util1}
\Delta_{a_i}u_i(S,X)
 & = & v(X_{i})
+ \underbrace{\sum_{j\neq i} a_{j} h}
               _\text{aggr. externalities} 
+ \underbrace{\phi \sum_{j\neq i} g_{ij} g_{ji} a_{j} }
               _\text{local externalities}
\end{eqnarray}
Note that $\Delta_{a_i}u_i(S,X)$ depends on $i$'s observables, her friendship links, and the choices of the overall population. The first term captures the possibility that the intrinsic preferences over different action statuses may depend on an agent's attributes - $v(X_{i})$, e.g., students in higher grades smoke more. The next two (summation) terms capture externalities. The local externalities terms $a_{j} g_{ij} g_{ji}\phi$ capture possible conformity pressures.\footnote{This is an example of positive externality. Alternative the local externality term may capture competitive pressures, in which case the term may have negative sign.} In the case of friendships, one may be influenced strongly by the behavior of own friends as opposed to casual individuals. On the other hand, $h$ in the second summation captures the aggregate externalities. A person may be potentially influenced from observing the behavior of the surrounding population, irrespectively of whether these are friends or not. In principle, $h$ could be a function on individual's exogenous attributes capturing, for example, a situation where males are more likely to be affected by the observed behavior of other males as opposed to the observed behavior of females.

Note how in formulation (\ref{util1}) there are two type of relationships: asymmetric, when there is a single link between two agents, and symmetric (reciprocal). In addition, for reasons that will become clear shortly, the model postulates that the externalities operate only through reciprocal relations. It is important to realize that these assumptions may (and should) be explicitly reasoned for given the context and the questions of a particular application. For example, in our data only around $35\%$ of the links are reciprocal. More interestingly, direct inspection (via correlation coefficients and OLS regression) of the dependence between individuals' behaviors and that of their non-reciprocal and reciprocal friends reveals that only the latter mediate statistically significant dependences. Given these observations, the model seems particularly suitable to study peer effects in teenage friendship networks.
Turning to the incremental payoff from a new link $\Delta_{g_{ij}}u_i(S,X)$ given as $u_i(g_{ij}=1,S_{-ij},X) - u_i(g_{ij}=0,S_{-ij},X)$, from (\ref{payoff}):\footnote{Again, the decisions do not need to be sequential. Rather I am presenting the \emph{incremental} payoffs of a friendship so the associations between functional forms and the object of interest are apparent.}
\begin{eqnarray} \label{util2}
\Delta_{g_{ij}}u_i(S,X)
& = & w(X_i,X_j)
+ \underbrace{g_{ji}m }_\text{reciprocity} 
+ \underbrace{a_{i}a_{j}g_{ji}\phi _{ij}}
       _\text{choice segregation}
+ \underbrace{q \sum_k g_{ik}g_{kj}}_\text{clustering/rivalry}
\end{eqnarray}
The first term $w(X_i,X_j)$ captures the baseline benefit from $i$'s unilateral decision to establish a link to $j$, which may or may not depend on their degree of similarity, i.e., same sex, gender, race, etc. The term $g_{ji}m\left( X_{i},X_{j}\right)$ shapes the degree of reciprocity.\footnote{The reciprocity varies substantially between applications. In our data from high school friendships for example the degree of reciprocity is below 40 percent. In financial network data, it may be as low as 10 percent.} The third term deserves special attention. It reflects the degree of similarity in the choices of $i$ and $j$, which is allowed to create addition stimuli for $i$ to establish a relationship with $j$. The last term captures the possibility of a link externalities. Mechanically, if $j$ links to $k$ and $k$ links to $i$ then $i$ may be more likely link to $j$ (thus closing the triangle). On the contrary, if there is friendship rivalry this logic will have the opposite effect, i.e. $q$ will be negative.


\comments{
Corporations which rely on banks to make payments or lend through the inter-bank payment network is yet another example. At a given point in time, the outcome $S$ is completely determined by the state of all edges $g_{ij}$ and vertexes $a_{i}$, i.e. $S=\left( \{ g_{ij}\} _{i,j=1}^n,\{ a_{i}\}_{i=1}^n\right)$.\footnote{We adopt the convention that $g_{ii}=0$.}

Let $\Gn$ be the set of all possible (directed and unweighted) $n$-node networks. Note that, even for small networks, $\Gn$ is a large set -- $\left\vert \Gn \right\vert =2^{n^2 }$ -- and grows exponentially with $n$.\footnote{For example, for $n=10$ , $ \left\vert \Gn \right\vert =2^{100}$.} Finally, it will be convenient to denote the state of the network including all edges and vertices but one edge with $S_{-ij}$. Similarly $S_{-i}$ fixes the network state excluding vertex (the action of agent) $i$.

\comments{
In particular, if $i$ nominates $j$ as a friend, then $g_{ij}=1$ ($i$ forms a one--directional link to $j$). Clearly, this notation does not preclude the existence of asymmetric relationships, i.e., individual $i$ may nominate $j$ as a friend but not vice versa.\footnote{One can argue that human friendships are symmetric by nature and should be modeled as such. However, I am interested in friendship relations as a mean to transmit influence, i.e., I am thinking about friends as role models. Moreover, only about $35\%$ of the friendships are reciprocal in my estimation sample.}

Figure \ref{diag1} illustrates the setup with $n=5$ agents. In the graph, each individual in $I=\{1,2,3,4,5\}$ is depicted as vertex and a friendship nomination is shown as a directed edge. In the example, the relationship between individuals $1$ and $2$ is symmetric, i.e., $g_{12}=g_{21}=1$. However, this is not the case for individuals $2$ and $3$ -- individual $2$ has nominated $3$ but the converse is not true ($g_{23}=1$ and $g_{32}=0$). Finally, in Figure \ref{diag1} the star-shaped nodes denote individuals with an action status equal to $1$ ($a_{3}=a_4=1$).\footnote{The model in this paper does not rely on any specific assumptions about the type of actions, other than being a binary choice that is potentially related to the friendship network.}
\begin{figure}[t]
\begin{center}
\caption{Example of Network State}
\vspace{0.5cm}
\def\node_index_color{\scriptsize\sffamily\bfseries\color{black}}
\def\linkfont{\color{white} \scriptsize}
\begin{tikzpicture}[->,scale=1.1,auto,every node/.style={scale=1.1},node distance=3cm]
\foreach \place/\name in {{(-2,0)/1},{(-0.8,1.4)/2},{(-0.8,-1.4)/5}}
\node[node] (\name) at \place   {\node_index_color\name};
\node[inf node] (3) at (2,0.7)  {\node_index_color 3};
\node[inf node] (4) at (2,-0.7) {\node_index_color 4};

\draw[transform canvas={xshift=-0.2ex,yshift= 0.2ex},->] (1) -- (2) ;
\draw[transform canvas={xshift=+0.2ex,yshift=-0.2ex},->] (2) -- (1) ;
\draw[transform canvas={xshift=+0.2ex,yshift=-0.4ex},->] (2) -- (3) ;
\draw[transform canvas={xshift=+0.0ex,yshift=-0.0ex},->] (2) -- (5) ;
\draw[transform canvas={xshift=+0.2ex,yshift= 0.2ex},->] (1) -- (5) ;
\draw[transform canvas={xshift=-0.2ex,yshift=-0.2ex},->] (5) -- (1) ;
\draw[transform canvas={xshift=-0.3ex,yshift= 0.0ex},->] (3) -- (4) ;
\draw[transform canvas={xshift=+0.3ex,yshift=-0.0ex},->] (4) -- (3) ;
\end{tikzpicture}
\label{diag1}
\end{center}
\fignotetitle{Note:} \fignotetext{Example with $n=5$ individuals and $8$ nominations. In $3$ cases the nominations are reciprocal - $g_{12}=g_{21}=1$, $g_{15}=g_{51}=1$, and $g_{34}=g_{43}=1$. The circle-shaped clear node indicates an action status of $0$, i.e., $a_1=a_2=a_5=0$, while the star-shaped shaded node indicates an action status of $1$, i.e., $a_3=a_4=1$.}
\end{figure}
}

Thus each agent $i$ selects her link nominations and action statuses $S_{(i)} = (a_i,\{g_{ij}\}_{j\neq i})$ from her choice set $\Si$. Each agent is endowed with a set of exogenous characteristics which together with her endogenous characteristics, e.g. network position and action status, shape her payoff

utility $u_i: \Sn \times \Xn\longrightarrow \R$, which maps the state $S\in \Sn$ and the attributes of the population $X=(X_1,...,X_n)\in \Xn$ to $\R$. Importantly, individuals' decisions impose externalities which propagate through the network of relationships. In various contexts these externalities may be referred to as peer effects or spillovers, and in their simplest form these are captured by the following payoff function:
\begin{eqnarray} \label{payoff}
u_i (S,X) & = & a_i v(X_i) + a_i \sum_j a_j h(X_i,X_j) + a_i\sum_j g_{ij}g_{ji} a_j \phi(X_i,X_j)\\ \notag
&   & + \sum_j g_{ij} w(X_i,X_j) + \sum_j g_{ij}g_{ji} m(X_i,X_j) + \gamma(X_i)\left( \sum_j g_{ij} \right)^2
\end{eqnarray}
Here $v(.)$, $\phi(.,.)$, $h(.,.)$, $w(.,.)$, and $m(.,.)$ are functions of agents' (exogenous) characteristics. The incentives encoded in ($\ref{payoff}$) are easier to understand, if we consider the incremental payoffs from changing just one dimension of an individual's decision, i.e., the incremental payoff of changing the action status or a single link.
}

\subsection{\emph{k}-player Nash stable network states and Nash partitions}
\comments{
\begin{enumerate}
\item Lemma links kPS and partition nash stable \& ordered by inclusion
\item Proposition existence
\item Prop non-perturbed dynamic in appendix, or just observation
\item Theorem k-player perturbed dynamic, convergence and closed form
\item Theorem k speed of convergence
\item Proposition link kNS and likelihood
\item Proposition link with QRE
\end{enumerate}
}

If agents make choices in their best interest, which patterns of links and behaviors are likely to prevail? Given agent's incentives, her observed links and behaviors are likely to compare favorably against possible alternatives, i.e. are likely to be robust against a set of feasible deviations. General considerations related to the complexity of individuals' decision problem could lend support to restricting the set of permissible deviations to local ones.\footnote{Indeed, the seminal paper of \cite{Jackson_Wolinsky_1996} proposes a notion of stability where the presence of a link implies that none of the parties involved has unilateral incentive to sever this link in isolation. The parties do not consider deviations involving severing multiple links at the same time.} Reasoning about the ``diameter'' of these local deviations, however, has not fallen in the focus of extant research and motivates a class of (non-cooperative) equilibria indexed by the diameter of the permissible local deviations. For $1 < k \le n$, $k$-player Nash stability fills in the gap between the local stability argument of \cite{Jackson_Wolinsky_1996} and the Nash play discussed by \cite{Bala_Goyal_2000}. 



A network state is $k$-player Nash stable, for $1 < k \le n$, if any set of $k$ agents play a Nash equilibrium of the induced game between the $k$ of them, conditioning on the rest of the network. In the induced game each participant chooses actions and friendship links (only to the participants) simultaneously. This formalizes the intuition that whenever individuals decide whether a ``profitable'' deviation is available, they do not consider all possible friendship combinations (which even for a population of size $100$ is $2^{99}$). Instead, individuals consider altering only part of their friendships at a time.

Let $\Gamma$ be the $n$-player normal form game $\{\Si,u_i\}_{i \in I}$ where as before $\Si$ is the choice set (strategy space) of agent $i$.
\begin{definition}
Let $S \in \Sn$ and $I_k = \{i_1,i_2,...,i_k\} \subseteq I$, for $1<k\le n$. Define the \textbf{\emph{restriction of $\Gamma$ to $I_k$ at $S$}} as the $k$-player normal form game $\Gamma |_{I_k}$, in the following way:
(i) The set of players is $\{i_1,i_2,...,i_k\}$;
(ii) The strategy space is $\mathbf{S}_{(i|I_k)} = \{0,1\}^k$, for $i \in I_k$, with typical element $S_{(i|I_k)} = \left(a_i,\{g_{ij}\}_{j\in I_k,j\neq i} \right)$;
(iii) The payoffs are given by the restriction $u_i|_{\mathbf{S}_{(I_k)}}$ of $u_i$ in (\ref{payoff}) on
$\mathbf{S}_{(I_k)}=\times_{i\in I_k} \mathbf{S}_{(i|I_k)}$.
\end{definition}

\begin{definition} For $1<k\le n$, the network state $S\in \Sn$ is said to be \textbf{k\emph{-player Nash stable}} provided that for any set of $k$ players $I_k\subseteq I$, the restriction of $S$ on $I_k$
\begin{equation*}
    S|_{I_k} = \left( \{a_i\}_{i\in I_k},\{g_{ij}\}_{i,j\in I_k} \right) \subseteq S
\end{equation*}
is a Nash equilibrium of $\Gamma |_{I_k}$ at $S$.
\end{definition}

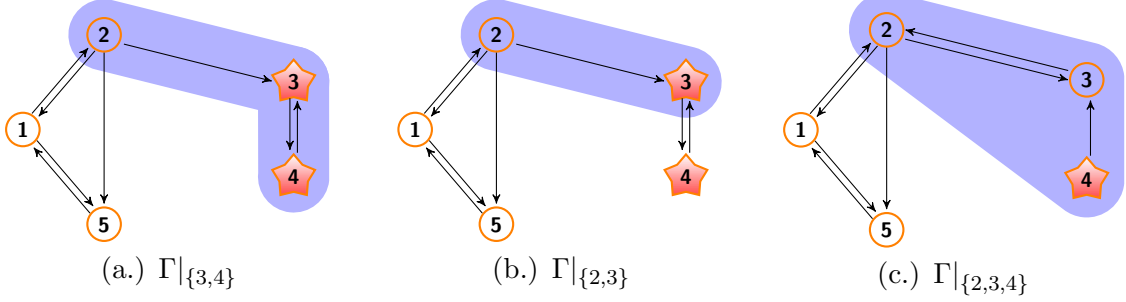
\begin{figure}
\caption{Example $2$- and $3$- player Nash stability} \label{diag2}
\begin{minipage}[c]{0.3\textwidth}
\centering
\def\node_index_color{\scriptsize\sffamily\bfseries\color{black}}
\def\linkfont{\color{black} \scriptsize}
\begin{tikzpicture}[->,>=stealth',shorten >=1pt,scale=0.9,auto,swap,node distance=3cm]
\foreach \place/\name in {{(-2,0)/1},{(-0.8,1.4)/2},{(-0.8,-1.4)/5}}
\node[node] (\name) at \place   {\node_index_color\name};
\node[inf node] (3) at (2,0.7)  {\node_index_color 3};
\node[inf node] (4) at (2,-0.7) {\node_index_color 4};

\draw[transform canvas={xshift=-0.2ex,yshift= 0.2ex},->] (1) -- (2) ;
\draw[transform canvas={xshift=+0.2ex,yshift=-0.2ex},->] (2) -- (1) ;
\draw[transform canvas={xshift=+0.2ex,yshift=-0.4ex},->] (2) -- (3) ;
\draw[transform canvas={xshift=+0.0ex,yshift=-0.0ex},->] (2) -- (5) ;
\draw[transform canvas={xshift=+0.2ex,yshift= 0.2ex},->] (1) -- (5) ;
\draw[transform canvas={xshift=-0.2ex,yshift=-0.2ex},->] (5) -- (1) ;
\draw[transform canvas={xshift=-0.3ex,yshift= 0.0ex},->] (3) -- (4) ;
\draw[transform canvas={xshift=+0.3ex,yshift=-0.0ex},->] (4) -- (3) ;
\begin{pgfonlayer}{background}
\fill[blue!30] \convexpath{3,2}{15pt};
\fill[blue!30] \convexpath{3,4}{15pt};
\end{pgfonlayer}
\end{tikzpicture}
(a.) $\Gamma|_{\{3,4\}}$
\end{minipage}
 \begin{minipage}[c]{0.02\textwidth}
 \hspace{5.0cm}
 \end{minipage}
\begin{minipage}[c]{0.3\textwidth}
\centering
\def\node_index_color{\scriptsize\sffamily\bfseries\color{black}}
\def\linkfont{\color{black} \scriptsize}
\begin{tikzpicture}[->,>=stealth',shorten >=1pt,scale=0.9,auto,swap,node distance=3cm]
\foreach \place/\name in {{(-2,0)/1},{(-0.8,1.4)/2},{(-0.8,-1.4)/5}}
\node[node] (\name) at \place   {\node_index_color\name};
\node[inf node] (3) at (2,0.7)  {\node_index_color 3};
\node[inf node] (4) at (2,-0.7) {\node_index_color 4};

\draw[transform canvas={xshift=-0.2ex,yshift= 0.2ex},->] (1) -- (2) ;
\draw[transform canvas={xshift=+0.2ex,yshift=-0.2ex},->] (2) -- (1) ;
\draw[transform canvas={xshift=+0.2ex,yshift=-0.4ex},->] (2) -- (3) ;
\draw[transform canvas={xshift=+0.0ex,yshift=-0.0ex},->] (2) -- (5) ;
\draw[transform canvas={xshift=+0.2ex,yshift= 0.2ex},->] (1) -- (5) ;
\draw[transform canvas={xshift=-0.2ex,yshift=-0.2ex},->] (5) -- (1) ;
\draw[transform canvas={xshift=-0.3ex,yshift= 0.0ex},->] (3) -- (4) ;
\draw[transform canvas={xshift=+0.3ex,yshift=-0.0ex},->] (4) -- (3) ;
\begin{pgfonlayer}{background}
\fill[blue!30] \convexpath{3,2}{15pt};
\end{pgfonlayer}
\end{tikzpicture}
(b.) $\Gamma|_{\{2,3\}}$
\end{minipage}
 \begin{minipage}[c]{0.02\textwidth}
 \hspace{5.0cm}
 \end{minipage}
\begin{minipage}[c]{0.3\textwidth}
\begin{center}
\def\node_index_color{\scriptsize\sffamily\bfseries\color{black}}
\def\linkfont{\color{black} \scriptsize}
\begin{tikzpicture}[->,>=stealth',shorten >=1pt,scale=0.95,auto,swap,node distance=3cm]
\foreach \place/\name in {{(-2,0)/1},{(-0.8,1.4)/2},{(-0.8,-1.4)/5}}
\node[node] (\name) at \place   {\node_index_color\name};
\node[node]     (3) at (2,0.7)  {\node_index_color 3};
\node[inf node] (4) at (2,-0.7) {\node_index_color 4};

\draw[transform canvas={xshift=-0.2ex,yshift= 0.2ex},->] (1) -- (2) ;
\draw[transform canvas={xshift=+0.2ex,yshift=-0.2ex},->] (2) -- (1) ;
\draw[transform canvas={xshift=+0.1ex,yshift=-0.4ex},->] (2) -- (3) ;
\draw[transform canvas={xshift=+0.0ex,yshift=-0.0ex},->] (2) -- (5) ;
\draw[transform canvas={xshift=+0.2ex,yshift= 0.2ex},->] (1) -- (5) ;
\draw[transform canvas={xshift=-0.2ex,yshift=-0.2ex},->] (5) -- (1) ;
\draw[transform canvas={xshift=-0.1ex,yshift=+0.4ex},->] (3) -- (2) ;
\draw[transform canvas={xshift=+0.3ex,yshift=-0.0ex},->] (4) -- (3) ;
\begin{pgfonlayer}{background}
\fill[blue!30] \convexpath{2,3,4}{15pt};
\end{pgfonlayer}
\end{tikzpicture}
(c.) $\Gamma|_{\{2,3,4\}}$
\end{center}
\end{minipage}
\end{figure}

Figure \ref{diag2} illustrates the notion of $k$-player Nash stability in a network with $n=5$ individuals. The left two diagrams show a network state which is $2$-player Nash stable. In such a state for every pair, the individuals within the pair are forming relationships with each other optimally (i.e., each individual plays best response to the strategy of the other). For example Diagram $(a.)$ considers the pair consisting of individuals 3 and 4. In the game between them $\Gamma|_{\{3,4\}}$, whenever individual 3 chooses $(a_3,g_{34})$ and individual $4$ chooses $(a_4,g_{43})$, the shaded outcome constitutes a Nash equilibrium. Similarly the shaded region in Diagram $(b.)$ depicts a Nash equilibrium of $\Gamma|_{\{2,3\}}$. Now suppose that although individual 3 best response in $\Gamma|_{\{2,3\}}$ and $\Gamma|_{\{3,4\}}$ is consistent with the depicted state, his best response in $\Gamma|_{\{2,3,4\}}$ would be to sever his friendship with 4, nominate 2 as a friend, and choose $a_3=0$. Diagram $(c.)$ depicts such a deviation which is not permissible in the context of $2$-player Nash stability.

Denote the set of all $k$-player Nash stable states with $\KNS$. Intuitively, as $k$ increases, $k$-player Nash stability places stronger requirements on what outcomes players consider as optimal (in the sense of being a best repose to the play of the rest). Indeed, the larger $k$ is the larger is the consideration set over which a player determines her best response. Consequently, one would expect $\KNS$ to shrink as $k$ increases. Below we formalize this intuition and provide an equivalent description of $k$-PS. 

\begin{definition} Consider a partition $\iota_m=\{I_{1},I_{2},...,I_{m}\}$ of the set of players $I$ i.e. $I=\cup I_j$ and $I_j \cap I_k =\emptyset$ for $j\ne k$. For the normal form game $\Gamma$, define a state $S \in \Sn$ to be \textbf{\emph{partition Nash stable with respect to partition}} $\iota_m$ provided $S|_{I_k}$ is a Nash equilibrium of $\Gamma |_{I_k}$ for all $I_k\in \iota_m$.
\end{definition}


\begin{lemma} \label{lemma:equilibrium_equivalence}
Let $\Gamma$ be the n-player normal form game $\{\Si,u_i\}$. Then the following holds:
\begin{enumerate}
 \item[(i)] A state $S\in \Sn$ is $k$-player Nash stable if and only if $S$ is partition Nash stable with respect to all partitions $\iota_m=\{I_{1},I_{2},...,I_{m}\}$ such that $|I_r|\leq k, \forall 1\leq r \leq m$.
 \item[(ii)] The family of $k$-player Nash stable states is ordered as
    $\TWONS\supseteq \ldots \KNS \supseteq \ldots \NNS.$
\end{enumerate}
\end{lemma}

In terms of equilibrium existence (and simple adaptive dynamic), our model falls in the category of potential games.\footnote{Congestion games were the first class of games exhibiting this property (\citealp*{beckmann_mcguire_winston_56,rosenthal_73}). \cite{monderer_shapley_96} recognize that congestion games are instances of games with potential, propose several notions of potential functions for games in strategic form, and obtain a characterization of potential games.} 

\begin{proposition}\label{prop:existence}
\textsc{[Existence]}
For any $1< k \le n$, there exists at least one $k$-payer Nash stable state.
Moreover the set of all $k$-player Nash stable states is characterized by:
\begin{equation}
 \KNS = \left\{ S \in \Gn : S_{(i|I_r)} \in \arg \max_{\tilde{S}_{(i|I_r)}} \Pfn(\tilde{S}_{(i|I_r)},S_{-(i|I_r)},X)  \quad \forall I_r\subseteq I, |I_r|\leq k \right\}
\end{equation}
where $\Pfn : \Gn \times \mathbf{X}_n \longrightarrow \mathbb{R}$ is a potential for the game in normal form.
\end{proposition}

\subsection{Adaptive behavior and a random utility framework}

In his doctoral thesis \cite{Nash1950} suggested that equilibrium might arise from some ``mass action'' as opposed to from a complex reasoning process.

\begin{quote}
We shall now take up the ``mass-action'' interpretation of equilibrium points. In this interpretation solutions have no great significance. It is unnecessary to assume that the participants have full knowledge of the total structure of the game, or the ability and inclination to go through any complex reasoning process. But the participants are supposed to accumulate empirical information on the relative advantages of the various pure strategies at their disposal.
\end{quote}

Indeed $k$-NS, including the Nash equilibrium, dispense with any behavioral model of network formation. While there is some argument for the state to persist once it is reached, it is not clear what behavioral mechanism will drive the system to such a state. Below I consider a family of myopic dynamic processes, which abstracts from any coordination between players and for which these stable states are both attractors and absorbing states.

\subsubsection{$k$-player dynamic ($k$-PD)}

The \emph{$k$-player dynamics} ($k$-PD) is a family of myopic dynamic processes which not only can give rise to an equilibrium as opposed to a complex reasoning process but also delivers a probabilistic ranking of $k$-PS. Consider a time dimension of the play, where individuals make their choices sequentially and revise these choices frequently.\footnote{For similar dynamic with $k=2$ see \cite{Cournot_1838,Nash1950,Blume_1993,JacksonWatts2002}.} In each period, which need not be of equal length, one agent adapts her strategy with respect to the current state of the network.\footnote{Note that, in rationalizing the individual's responses, such models put all the weight on the contemporaneous environment rather than on the human ability to forecast the actions of the other. This is an adequate visualization of the reality under at least two scenarios. First, if the action of a single individual cannot substantially affect the future development of the network. In such cases individuals do not need to behave strategically with respect to the future response of the network. Second, if individuals form stationary expectations about the future that are, in fact, consistent with the current network state. In this case, although individuals need to act strategically with respect to the future, their best response to the current state is indeed a best response to the expected network state tomorrow as well. Thus, the proposed approach is likely to be plausible in studying decisions such as smoking and drinking, while it seems inappropriate in situations where an individual's forecast about the future is likely to be important (for example, to study fertility or career choices).} %
More specifically, a randomly chosen individual (say $i$) meets a random set $I_{k-1}$ of $k-1$ individuals from the population $I$. All individuals in $I_{k-1}$ are potential friends and upon meeting them $i$ decides jointly whether to befriend any (or all) of them and whether or not to change her action status $a_i$. The rules of the play in period $t$ are determined by the outcome of a (stochastic) meeting process $\mu_t$, which outputs who makes choices and who are the individuals considered as potential friends. Formally:
\begin{equation} \label{meeting}
\Pr \left( \mu _{t}=\left( i,I_k\right) |S_{t-1},X\right)
= \mu_{i,I_k} \left(S_{t-1},X\right)
\end{equation}%
It is clear that the sequence of meetings and players' optimal decisions induce a sequence of network states $(S_t)$, which is indexed by time subscript $t$.\footnote{In the simplest case, any meeting is equally probable so that: $\mu_{i,I_k} \left(S_{t-1},X\right)=\frac{1}{n} \frac{1}{{n-1 \choose k}}$. Also more general meeting technologies are possible, e.g. with bias for people with similar actions and/or characteristics, as, for example, $\mu_{i,I_k} \left(S_{t-1},X\right) \propto \exp \left\{-\max_{j\in I_k}\{d(a_i,a_j)\}-\max_{j\in I_k}\{d(X_i,X_j)\}\right\}$ with $d$ being a generic distance function.}
 
Note that each $k$-PS is a \emph{rest} point of a meeting process of dimension $k$---loosely speeking, if the network reaches such a state, it will stay there forever. In addition, minimal assumptions guarantee that $k$-PS are \emph{attractors} of the proposed dynamic. 

\begin{assumption} \label{asm:meet1}
Any meeting is possible 
$\Pr \left( \mu_t =(i,I_{k-1}) | S,X \right) > 0 $
for all $i \in I$, $I_k\ne \emptyset$, $S \in \Gn$ and $X \in \Xn$.
\end{assumption}


\begin{proposition} \label{prop:dynamics}
Let $1<k\le n$ and suppose that assumption \ref{asm:meet1} holds then
\begin{enumerate}
  \item Any $S\in \KNS$ is absorbing i.e. if $S_t \in \KNS$ then $S_{t'}=S_t$ for all $t'>t$.
  \item Independently of the initial condition (distribution)
 $\Pr \left( \lim_{t\rightarrow \infty} S_t \in \KNS \right) = 1.$
\end{enumerate}
\end{proposition}

The first part of proposition \ref{prop:dynamics} is implied immediately from the definition of $\KNS$. In this setting, the second part follows elegantly from the observation that the potential function $\{\Pfn_t \}$ is a sub-martingale.

While the dynamic perspective introduced above addresses the concern raised by \citet*{kandori_mailath_rob_93}, namely, it explains how equilibrium is reached from the behavior of the agents, it exhibits some undesirable properties. First and foremost, it does not provide a convenient statistical framework that can be estimated given network data. In the model, the uncertainty is driven solely by the meeting process, which is likely to be unobservable. Thus, which equilibrium will be played is entirely determined by the realization of $\mu$. Moreover, it is not \emph{a priori} clear which mode of the meeting process is observed in the data, i.e., what is $k$, and different $k$ will result in observing different classes of equilibria. 



\subsubsection{A random utility framework}

The assumption below introduces to this discrete choice problem a random preference shock very much in the spirit of a random utility model.\footnote{See \cite{thurstone_27,marschak_60,mcfadden_74}.} Similar stochastic has been considered by the literature on stochastic stability, which when shocks vanish over time, presents an equilibrium selection device.\footnote{See \cite{foster_young_90,kandori_mailath_rob_93}.}

\begin{assumption} \label{asm:error_utility}
Suppose that the utilities in (\ref{payoff}) contain a random preference shock. More specifically, let
\begin{equation*}
  \overline{u}_i(S,X)=u_i(S,X)+\epsilon_{i,S}
\end{equation*}
with $\epsilon_{i,S} \sim i.i.d.$ across time, individuals and network states. Moreover, suppose that $\epsilon$ has c.d.f. and unbounded support on $\mathbb{R}$.
\end{assumption}

\begin{assumption} \label{asm:gumbel}
Suppose that the preference shock $\epsilon$ is distributed $Gumbel (\underline{\mu},\underline{\beta})$.
\end{assumption}

\begin{assumption} \label{asm:meet2}
 Suppose that for the meeting probability $\mu$: (i) $\Pr (\mu_t = (i,I_k))$ does not depend on the relationship status between $i$ and any $j\in I_k$. (ii) $\Pr (\mu_t = (i,I_k))$ does not depend on $a_i$. These together imply $\mu_{i,I_k}(S,X) = \mu_{i,I_k}(S',X)$ for all $S=(S_{(i|I_k)},S_{-(i|I_k)})$ and $S'=(S'_{(i|I_k)},S_{-(i|I_k)})$.
\end{assumption}

The matching process $\{\mu_t\}_{t=1}^{\infty}$ and the sequence of optimal choices, in terms of friends selection and individual actions, induce a Markov chain of network configurations on $\Gn$.\footnote{See the stochastic-choice dynamics in \cite{Blume_1993}.} The above assumption guarantees that this chain obeys some desirable properties, which are formalized below. First, I lay out a formal statement of a theorem and then discuss its implications. (The proof is in the appendix on p. \pageref{proof:thm:stationarydistribution}.)

\begin{theorem} \label{thm:stationarydistribution}
\textsc{[Stationary distribution]}
Let $1<k\le n$ and suppose assumptions \ref{asm:meet1} and \ref{asm:error_utility} hold. The Markov chain generated by the $k$-PD has the following properties:
\begin{enumerate}
 \item There exists a unique stationary distribution $\pi_k \in \Delta(\Gn)$ for which $\lim_{t \rightarrow \infty} \Pr (S_t = S) = \pi_k(S)$. In addition, for any function $f:\Gn \rightarrow \mathbb{R}$, 
$\frac{1}{T} \sum_{t=0}^T f(S_t) \longrightarrow \int f\left( S\right) d \pi_k.$

\item Under assumptions \ref{asm:meet1}-\ref{asm:meet2}:
\begin{equation}\label{logl}
 \pi_k(S,X) \propto \exp \left( \frac{\Pfn(S,X)}{\underline{\beta}} \right)
\end{equation}
In particular, $\pi_k(S,X)$ does not depend on $k$.
\end{enumerate}
\end{theorem}

The first part is not surprising in that it asserts that the Markov chain, generated by the $k$-player meeting process, is well behaved so that standard convergence results apply. The uniqueness of $\pi_k$ precludes any ambiguity in the predictions of the process, while the ergodicity is relevant whenever we want to draw predictions from the model. However, the second part of the theorem has novel implications.

Note how in (\ref{thm:stationarydistribution}) the stationary distribution does not depend on $k$ and, thus, delivers an approach to unify the family of $k$-PS. As we formally argue in theorem \ref{thm:equlibriumranking}, the stationary distribution offers a probabilistic ranking of the set $k$-player stable states (within and across different $k$s). In addition, the closed-form expression for the stationary distribution has advantages for the empirical implementation of the proposed framework, where $\pi_k$ can be treated as the likelihood. In particular, one can explore a transparent argument for the identification of model's parameters. It is clear that, given the variation in the data of individual choices $\{a_i\}_{i=1}^n$, friendships $\{ g_{ij} \}_{i,j=1}^n$ and attributes $\{ X_i \}_{i=1}^n$, functional forms for $v,w,m,h,q$ will be identified as long as the different parameters induce different likelihoods of the data. Also, a closed-form expression for $\pi_k$ facilitates the use of likelihood-based estimation methods, including Bayesian ones.\footnote{To provide intuition behind this result consider two states $S,S' \in \Sn$. It can be shown that the probability of moving from $S$ to $S'$ is proportional to the probability of returning from $S'$ to $S$ by a factor that does not depend on $k$. The formal argument can be found in the appendix.}

The second main question in this section is how to differentiate the family of $k$-player dynamic with respect to how well, for a given $k$, observations from this dynamic is represented by the stationary distribution $\pi_k$. We obtain a sharp characterization with respect to the second eigen value of the transient operator of the $k$-PD.\footnote{In our case the speed of convergence is shaped by how powers of the transient matrix map any initial state to the eigenvector with eigenvalue one. All eigen values except one are strictly less than $1$ so that the limiting behavior of this exponentiation is governed by the second eigenvalue.} The assertion is formulated in the most stark case when the only factor influencing the speed of convergence is the dimension of the meeting process.\footnote{In general, the shape of the potential and the geography of the network will likely influence the speed of convergence. For more see \cite{BhamidiBrestlerSly_2011}. As far as I know treatment of the most general case remains out of reach.}
\begin{theorem}\label{thm:convergencespeed}
\textsc{[$k$-PD ranking]}
Suppose that $\Pfn(S)=c$ for all $S\in \SSn$. For $2 \leq k < k' \leq n$, the $k'$-PD converges strictly faster than $k$-PD to the stationary equilibrium $\pi$. In particular, the second eigen value of $k$-PD is given by
\begin{equation}\label{eigen2}
\lambda_{k,[2]}=\frac{1}{n}\left(n-1+\frac{n-k}{n-1}\right)
\end{equation}
\end{theorem}

There are two rationales behind the pursuit of a characterization of the speed of convergence of $k$-PD. As anticipated (and formally established shortly) $\pi_k$ probabilistically ranks the family of $k$-PS and (of course) Nash equilibria are clear favorites. In a dual fashion, the differential speed of convergence provides a means to rank probabilistically the $k$-PD in the sense of how likely are these to be represented by the stationary distribution. Theorem \ref{thm:convergencespeed} suggests that for large $k$s, for which the unperturbed $k$-PD is more likely to converge to a Nash equilibrium, $k$-PD converges the fastest.

The second reason for why studying how well the $k$-PD is represented by the stationary distribution $\pi_k$ was highlighted first by \cite{BhamidiBrestlerSly_2011} who showed that ``mass action'' dynamics with local updates converges very slowly.\footnote{A Markov chain is local if at most $o(n)$ links are updated at a time.} Note that $k$-PD encompasses not only local updates, e.g. $k=[n/2]$, and thus offers a way to address the problem of slow convergence (poor approximation).\footnote{This question is relevant not only the positive analysis of the $k$-PD but, from a purely practical view point, also for the implementation of various estimation/simulation techniques when working with the model (\citealp{ArunJackson2016}).} 
In addition, theorem \ref{thm:convergencespeed} offers insights into an important trade-off motivated by the practical implementation of the framework. In designing sampling schemes the Markov chain is facing a trade-off between speed of convergence and complexity in simulating the next step. Whenever $k$ is small, the speed of convergence to the stationary distribution is slower, however, the computational difficulty in updating the network at each step is smaller. The opposite holds when $k$ is large.

\subsection{Discussion}

The family of stochastic best response dynamics (indexed by dimension $k$) has a particular meaning in our settings. Their (common) stationary distribution delivers a device to probabilistically \emph{rank} these equilibria. Below we return to this point, present a generalize meeting process (random-$k$-$k$-player), and briefly comment on the link with the notions of a quantal response equilibrium (\citealp{mckelvey_palfrey_95}) and a correlated equilibrium (\citealp{aumann_74}).

\subsubsection{Probabilistic ranking and long-run equilibria}

The stationary distribution obtained in theorem \ref{thm:stationarydistribution} gives an intuitive (probabilistic) ranking of the family of $k$-player Nash stable equilibria. Under $\pi$, all $k$-Nash stable states receive a positive probability (across all possible values of $k$). Moreover, and realistically, a network state will receive a positive probability, although it may not be an equilibrium in any sense. It will be desirable, however, that in the vicinity of an equilibrium, the equilibrium to receive the highest probability. Relatedly, the mode of $\pi$ (i.e. the state with the highest probability) has special role. The goal in this section is to provide a new perspective to the theoretical results on equilibrium selection from evolutionary game theory, namely equilibrium ranking.

To formalize our discussion, define the neighborhood $\Ns \subset \Gn$ of $S \in \Gn$ as:
\begin{equation*}
  \Ns = \left\lbrace S': S'=(g_{ij},S_{-ij}), i\neq j \right\rbrace \bigcup
  \left\lbrace S': S'=(a_{i},S_{-i}) \right\rbrace
\end{equation*}
Next, define a state $S$ as a \emph{long-run} equilibrium of the network formation model if for any sequence of vanishing preference shocks, the stationary distribution $\pi$ places a positive probability on $S$ (\citealp{kandori_mailath_rob_93}).

\begin{theorem} \label{thm:equlibriumranking}
Suppose assumptions \ref{asm:meet1}-\ref{asm:meet2} hold:
\begin{enumerate}
  \item A state $S\in \Sn$ is a $2$-player Nash stable if and only if it receives the highest probability in its neighborhood $\Ns$.
  \item The most likely network states $\Smode \in \Sn$ (the one where the network spends most of its time) are Nash equilibrium of the one-shot game i.e. $\Smode \in \NNS$.
  \item The long-run equilibria of the underlying evolutionary model that are given by $\Smode$, which need not be Pareto efficient.
\end{enumerate}
\end{theorem}

\subsubsection{Meeting process of random dimension}

Now consider what appears to be a very unrestrictive meeting process, where every period a random individual meets a set of potential friends of random size and composition. Let $\kappa$ be a discrete process with support $2,\ldots,n$ and augment the meeting process with an additional initialization step with respect to the dimension of $\mu$. In particular, at each period first $\kappa$ is realized and then $\mu^k$ is drawn just as before. It is relatively straightforward to establish, without any assumptions on the process $\kappa$, that this augmented process has the same stationary distribution $\pi$ as the one from theorem \ref{thm:stationarydistribution}.\footnote{I omit here the formal statement and the proof as it essentially follows the one from \ref{thm:stationarydistribution}.} This is another demonstration of the fact that different meeting processes result in observationally equivalent models. 

\subsubsection{Quantal response equilibrium}

The notion of a quantal response equilibrium (QRE), introduced by \cite{mckelvey_palfrey_95}, is a fixed point of the quantal-response functions (QRFs), very much like Nash equilibrium is a fixed point of the best response functions. The QRF of player $i$ is a smoothed best response function, where the strictly rational choice of player $i$ (i.e., the best response) is replaced by an approximately rational response. A (regular) quantal-response function satisfies the following axioms:
\begin{enumerate}[topsep=1ex,parsep=0ex,itemsep=0ex,leftmargin=0.7cm,label=(\roman*)]
 \item \bsc{Interiority:} every strategy in $i$'s strategy space receives a strictly positive probability.
 \item \bsc{Continuity:} the probability of player $i$ choosing pure strategy $s$ is a continuously differentiable function of $i$'s expected payoff from choosing $s$.
 \item \bsc{Responsiveness:} the first derivative of the above probability is strictly positive for all players on the support of the expected payoffs.
 \item \bsc{Monotonicity:} strategies with higher expected payoff receive a higher probability.
\end{enumerate}

The existence of a (regular) quantile response equilibrium of a finite-player finite-strategy space normal form game trivially follows from Brower's fixed point theorem. Any such equilibrium induces a probability distribution $\pi^{QRE}$ over $S=S_1 \times S_2 \times ... \times S_n$ where $S_i$ is the set of pure strategies of player $i$. Note that $\pi^{QRE}$ is the Cartesian product of the equilibrium quantile responses and whence inherits their properties - i.e., the conditional distributions satisfy the axioms above. By way of comparison, the stationary distribution $\pi$ of the network formation model bears some similarities to the axioms of QRF inherited in $\pi^{QRE}$. However, there are important differences. I discuss each in turn.

\begin{proposition} \label{prop:qre}
The conditional distribution of player $i$'s choices (i.e., $a_i,\{g_{ij}\}_{j\neq i}$) on the choices of the rest of the players -- $\pi_{(i)|(-i)}\in \Delta (\{0,1\}^n)$ -- induced by the optimal play in the network formation model satisfies all properties of a (regular) quantile response function - interiority, continuity, responsiveness, and monotonicity. However, a QRE cannot induce $\pi$.
\end{proposition}

The first part follows trivially from the expression for $\pi$, given by theorem \ref{thm:stationarydistribution}. Since the discrete network formation game is a game with potential, all payoffs can be represented by a single function of the network state - the potential $\Pfn$. Importantly, the potential ranks the network states consistently with individual preferences and, at the same time, is linked to the probability of observing a given state. Thus, it is not surprising that $\pi$ exhibits the intuitive properties of $\pi^{QRE}$.

For the second part, note that the sequential decision process induces a correlation between players' decisions. In particular, the distribution $\pi$ in theorem \ref{thm:stationarydistribution} cannot necessarily be factored as a product of marginal probability distributions over each player choice set $\mathbf{S}_{(i)}$. (for a formal example see page \pageref{proof:prop:qre})  It then follows that QRE cannot induce $\pi$. For the same reason, $\pi$ cannot be induced by a mixed strategy profile of the unperturbed $n$ player one-shot network formation game. In essence, the outcome of the sequential network play is bears similarities to the notion of a correlated equilibrium in \cite{aumann_74}.

\section{Data and estimation}
I implement the model with data on smoking and friendships from the National Longitudinal Study of Adolescent Health (Add Health) data. The estimation is non-standard in that the likelihood can be evaluated only up to an intractable constant. Fortunately, both frequentist and Bayesian methods relying on MCMC algorithms have been adapted to deal with these complications.\footnote{See \cite{geyer_thompson_92,murray2012mcmc,Mele2017}.} In addition, this particular implementation benefits the intuition of theorems \ref{thm:stationarydistribution} and \ref{thm:convergencespeed}, which have direct implications for constructing algorithms for estimation and simulation from the model.

\subsection{The Add Health data}
The National Longitudinal Study of Adolescent Health is a longitudinal study of a nationally representative sample of adolescents in grades $7$--$12$ in the United States in the $1994$--$95$ school year. In total, 80 high schools were selected together with their ``feeder`` schools. The sample is representative of US schools with respect to region of country, urbanicity, school size, school type, and ethnicity. The students were first surveyed in-school and then at home in four follow-up waves conducted in $1994$--$95$, $1996$, $2001$--$02$, and $2007$--$08$. This paper makes use of Wave I of the in-home interviews, which contain rich data on individual behaviors, home environment, and friendship networks.\footnote{In addition to the in-home interview from Wave I, data on friendship are available from the in-school and Wave III interviews. However, the in-school questionnaire itself does not provide information on important dimensions of an individual's socio-economic and home environment, such as student allowances, parental education, and parental smoking behaviors. On the other hand, during the collection of the Wave III data, the respondents were not in high school any more. For more details on Add Health research design, see \url{www.cpc.unc.edu/projects/addhealth/design}.}

To provide unbiased and complete coverage of the social network, \emph{all} enrolled students in the schools from the so-called saturated sample were eligible for in-home interviews. These were 16 schools of which 2 large schools (with a total combined enrollment exceeding 3,300) and 14 small schools (with enrollments of fewer than 300). One of the large schools is predominantly white and is located in a mid-sized town. The other is ethnically heterogeneous and is located in a major metropolitan area. The 14 small schools, some public and some private, are located in both rural and urban areas.

In addition, Add Health data have been merged with existing databases with information about respondents' neighborhoods and communities. For example, the American Chamber of Commerce Research Association (ACCRA) compiles cost of living index, which is linked to the Add Health data on the basis of state and county FIPS codes for the year in which the data were collected. From the ACCRA, I use administrative data on the average price of a carton of cigarettes.\footnote{For details see the Council for Community and Economic Research \url{www.c2er.org}, formerly the American Chamber of Commerce Research Association.} Additional details about the estimation sample including sample construction and sample statistics are presented in the appendix.

\subsection{Bayesian estimation}
Given the $k$-player meeting process, theorem \ref{thm:stationarydistribution} demonstrates that the network state evolves according to a Markov chain with a unique stationary distribution $\pi$ on the set of all network states $\Sn$. Because no information is available on when the network process started or on its initial state, the best prediction about the network state is given by $\pi$. Thus, for estimation purposes, the stationary distribution can be thought of as the likelihood. Given a single network observation $S \in \Sn$, the likelihood is given by:%
\begin{equation}
  p(S|\theta) = \frac{\exp\{\Pfn_{\theta}(S)\}}{H_{\theta}}
\end{equation}
where $\Pfn_\theta$ is the potential (evaluated at $\theta$) and $H_{\theta}=\sum_{S \in \Sn} \exp\{S\}$ is an (intractable) normalizing constant. The intractability of $H_{\theta}$ encapsulates the fact that the size of $\Sn$ and the summation in calculating $H_\theta$ are so large, even for small networks, that the value of $p(S|\theta)$ cannot be calculated directly for practical purposes.\footnote{For $n=10$ the summation includes $2^{100}$ terms.}

The specific form of the likelihood pertains to the exponential family, whose application to graphical models has been termed as Exponential Random Graph Models.\footnote{For more see \cite{FrankStrauss1986markov,WassermanPattison1996,Mele2017,ArunJackson2016}.} Various generative and descriptive approaches have been proposed, both within frequentist and Bayesian paradigms, to address this specific tractability problem we outlined above. This paper adopts a Bayesian approach and our algorithm for sampling from the posterior is an adaptation of the double M-H algorithm of \citealp{murray2012mcmc} and \cite{Mele2017}, informed by theorem \ref{thm:convergencespeed}.

The posterior sampling algorithm is exhibited in table \ref{table:algorithm}. In the original double M-H algorithm, an M-H sampling of $S$ from $\pi_\theta(S)$ is nested in an M-H sampling of $\theta$ from the posterior $p(\theta\S)$. The novelty of my approach is the random dimension of the meeting process in step $5$. Theorem \ref{thm:convergencespeed} suggests that varying $k$ improves the convergence and theorem \ref{thm:stationarydistribution} demonstrates that changing $k$ leaves the stationary distribution unaltered. The validity of the particular implementation is proven in proposition \ref{prop:varying_mh}.

\begin{table}[!t]
\caption{Varying double M-H algorithm}
\label{table:algorithm}
\begin{center}
Input: initial $\theta^{(0)}$, number of iterations $T$, size of the Monte Carlo $R$, data S \\
\begin{tabular}{cl}
\hline
1. & \textbf{for} $t=1\ldots T$ \\
2. & \quad Propose $\theta' \sim q(\theta';\theta^{(t-1)},S)$\\
3. & \quad Initialize $S^{(0)}=S$\\
4. & \quad \textbf{for} $r=1\ldots R$ \\
5. & \quad \quad Draw $k \sim p_k(k)$\\
6. & \quad \quad Draw a meeting $\mu(i,I_k)$ where $i\in \{1\ldots N\}$ and $I_k \subset \{1\ldots N\}\backslash\{i\}$ from $q_\mu(i,I_k)$\\
8. & \quad \quad Propose $S'$ where $(a_i,\{g_{ij}\}_{j\in I_k})$ are drawn from $q_\mu(S'|S^{(r-1)};(i,I_k))$ \\
9. & \quad \quad Compute 
$\bar{a} = \frac{\exp\{\Pfn_{\theta'}(S')\}}{\exp\{\Pfn_{\theta'}(S^{(r-1)})\}} 
           \frac{Q(S^{(r-1)}|S';p_k,q_{i,I_k})}{Q(S'|S^{(r-1)};p_k,q_{i,I_k})}$ \\
10. & \quad \quad Draw $a\sim \text{Uniform}[0,1]$\\
11. & \quad \quad If $a<\bar{a}$ then $S^{(r)}=S'$ else $S^{(r)}=S^{(r-1)}$\\
12. & \quad \textbf{end} for $[r]$ \\
13. & \quad Compute 
$\bar{a}=\frac{q(\theta';\theta^{(t-1)})}{q(\theta^{(t-1)};\theta')}
\frac{p(\theta')}{p(\theta^{(t-1)})}
\frac{\exp\{\Pfn_{\theta^{(t-1)}}(S^{(R)})\}}{\exp\{\Pfn_{\theta^{(t-1)}}(S)\}}
\frac{\exp\{\Pfn_{\theta'}(S)\}}{\exp\{\Pfn_{\theta'}(S^{(R)})\}}$\\
14. & \quad Draw $a\sim \text{Uniform}[0,1]$\\
15. & \quad If $a<\bar{a}$ then $\theta^{(t)}=\theta'$ else $\theta^{(t)}=\theta^{(t-1)}$\\
16. & \textbf{end} for $[t]$ \\ \hline
\end{tabular}
\end{center}
\end{table}

\begin{proposition}
\textsc{[Varying double M-H algorithm]}
\label{prop:varying_mh}
Let $1<k\le n$ and suppose assumptions \ref{asm:meet1} and \ref{asm:error_utility} hold. If in the algorithm of table \ref{table:algorithm}, the proposal density conditional on meeting $(i,I_k)$ of random dimension $k$, $q_\mu(S'|S);(i,I_k))$ is symmetric, then the unconditional proposal $Q(S'|S)$ is symmetric. In particular, the acceptance ratio of the inner M-H step 9 does not depend on $p_k$ and $q_\mu$.
\end{proposition}

Finally, the Bayesian estimator requires specifying prior distributions and proposal densities. The prior $p(\theta)$ is normal centered around the OLS coefficients of the respective linear regressions. The variances are chosen wide enough so that $0$ is a standard deviation away from the mean. Proposals $p_k$, $\mu$, and $q_\mu$ are uniform over their respective domains.
\subsection{Parametrization}
I explore a wide set of parametrizations which are informed by the salient features of the data and by the particular experiments I am interested in.\footnote{Importantly, the trade off between flexibility and data limitation is tight. Recall that our identification framework is casted in the many networks asymptotics and, in the end, we have 12 networks.} Careful scrutiny of the patterns of smoking by socioeconomic factors together with the mixing patterns of smokers and non-smokers motivates the following incremental payoffs given in (\ref{util1})
\begin{eqnarray*}
\Delta_{a_i}u_i(S,X)    & = & v(X_i)
 + h \sum_{j\neq i} a_{j} + \phi \sum_{j\neq i} a_{j} g_{ij} g_{ji} + \epsilon_{a,i}\\
\end{eqnarray*}
with baseline utility of smoking $v(X_i) = v_0 + v_1 \ln (y) + \sum_{.} v_{.}d_{.}(X_i)$ where the indicator functions $d_{.}(X_i)$ include: (i.) smoker in the household, (ii.) mother's education (high school or some college), and (iii.) dummy race (only for) blacks.

\comments{
I parametrize the contextual effect term $c(X_{i},X_{j})$ in a way that will enable the model to capture the observed regularity in the data that individuals involved in a cross-gender friendship where the male is older than the female are more likely to smoke. In particular $c(X_{i},X_{j})=\zeta (d_1(X_i,X_j)+d_2(X_i,X_j))$ where
\begin{center}
\begin{tabular}{ll}
 $d_1(X_i,X_j)$ & $=$ 1 if $sex_i=male$, $sex_j=female$, and $grade_i>grade_j$ \\
 $d_2(X_i,X_j)$ & $=$ 1 if $sex_i=female$, $sex_j=male$, and $grade_i<grade_j$
\end{tabular}
\end{center}
}

The incremental utility of adding a link 
given in (\ref{util2}) becomes:
\begin{eqnarray*}
\Delta_{g_{ij}}u_i(S,X) & = & w(X_i,X_j) + mg_{ji} + \phi g_{ji}a_{i}a_{j}  \\
                              &   & +2d_i +1 + \sum_k (g_{jk}+g_{kj})(g_{ki}+g_{ik}) q + \epsilon_{g,ij}
\end{eqnarray*}
the first term $w(X_i,X_j)$ contains dummies for: (i.) different sex, (ii.) different grade, (iii.) different race, and $d_i$ stands for the number of friends of $i$.\footnote{So that the marginal payoff of friendships is decreasing provided $d<0$.}

\subsection{Identification}
The identification, within the framework of many networks, follows trivially from the connection of the model with the family of exponential random graph models (ERGM). These are a broad class of statistical models, capable of incorporating arbitrary dependencies among the links of a network. As a result, ERGM have been very popular in estimating statistical models of network formation (see \citealp{jackson_08}). A powerful corollary of theorem \ref{thm:stationarydistribution} is that the likelihood of the model falls in the family of ERGM. In more general terms, it follows that ERGM are broad enough to incorporate the strategic incentives of the static one-shot play embedded in $\pi$.
\begin{corollary}
The likelihood of the model can be written as
$ l(\theta|S) \propto \exp\left\{\sum_{r=1}^R\theta_i w_i(S,X) \right\} $,
where $w_i:\Sn \times \Xn \longrightarrow \R$.
\end{corollary}

As the number of networks grows to infinity, identification follows from the theory of the exponential family. In particular, it is enough that the sufficient statistics $w_i$ are linearly independent functions on $\Sn \times \Xn$. In the structural parametrization of the model above, this condition is established immediately.\footnote{Most of the parameters are identified in the asymptotic frame, where the size of the network grows to infinity (as opposed to the number of networks going to infinity). Further discussion of this point is outside the scope of this work. For more see \cite{xu_11,ArunJackson2016}.}

\subsubsection*{Unobservable heterogeneity in friendship selection and decision to smoke}

In addition to the models' parameters for observable attributes, it is possible to incorporate agents' specific unobservable types $\tau_i \sim N(0,\sigma^2_{\tau})$ which may influence \emph{both} the utility for friendships, e.g. $W(.,.)$ could include term $|\tau_i-\tau_j|$, and also the propensity to smoke, e.g. $V(.)$ could include a term $\rho_{\tau}\tau_i$. In this case the likelihood has to integrate out $\vec{\tau}$:
\begin{equation}
p(S|\theta) = \int_{\vec{\tau}} \frac{\exp\{\Pfn_\theta(S,\vec{\tau}\}}{\sum_{\hat{S}} \exp\{\Pfn_\theta(\hat{S},\vec{\tau})\} } \phi(\vec{\tau}) d\vec{\tau}
\end{equation}

For this extension, in the many network asymptotics framework, the heuristic identification argument goes as follows. Individuals who are far away in observables, must have realizations of the unobservables very close by. If in the data those individuals are either smokers or non smokers with very high probability then it must be the case that $\rho_\tau$ is large. 

Adding unobservable heterogeneity imposes substantial computational costs to the estimation and simulation algorithms because these need to include an additional step---drawing from $\vec{\tau}$. For the Add Health data, such an extension is out of reach because of the limited number of well-sampled networks. 

\subsection{Estimation results}

Table \ref{tab:estimates} presents the posterior means and indicates whether the shortest $90\%$, $95\%$, and $99\%$ credible sets of the posterior sample contain zero. To facilitate the interpretation of the estimates table \ref{tab:estimates} reports the baseline and marginal probabilities directly. For example, the parameter on the baseline utility of smoking $\theta_1$ in fact equals $\frac{e^{v_0}}{1+e^{v_0}}$.\footnote{So that $v_0=\ln(\theta_1)-\ln(1-\theta_1)$.} The index MP stands for marginal probability and MP$\%$ stands for marginal probability in percentages (with respect to the baseline probability).

\begin{table}[t!]
\begin{center}
  \caption{Parameter Estimates} \label{tab:estimates}%
    \begin{tabular}{llccccc}
  \hline \hline
    \multicolumn{4}{c}{\textit{Utility of smoking}}       \\
          & Parameter                                & No Net Data    & Fixed Net         & Model I          & Model II         & Model III      \\ \hline
    1     & Baseline probability of smoking          & $0.198^{***}$   & $0.121^{***}$     & $0.129^{***}$    & $0.107^{***}$    & $0.124^{***}$  \\
    2     & Price                                    & $-0.002$        & $-0.002$          & $-0.001$         & $-0.001$         & $-0.003^{*}$    \\
    3     & HH smokes$^{MP}$                         & $0.091^{***}$   & $0.113^{***}$     & $0.109^{***}$    & $0.103^{***}$    & $0.115^{***}$   \\
    4     & Mom edu (HS\&CO)$^{MP}$                  & $-0.031^{***}$  & $-0.035^{***}$    & $-0.039^{***}$   & $-0.030^{***}$   & $-0.034^{***}$  \\
    5     & Blacks$^{MP}$                            & $-0.169^{***}$  & $-0.179^{***}$    & $-0.189^{***}$   & $-0.167^{***}$   & $-0.187^{***}$   \\
    6     & Grade 9+$^{MP}$                          & $0.128^{***}$   & $0.139^{***}$     & $0.125^{***}$    & $0.151^{***}$    & $0.139^{***}$   \\
    7     & $30\%$ of the school smokes$^{MP}$       & $0.079^{***}$   & $0.077^{***}$     & $0.084^{***}$    & $0.073^{***}$    & $0.074^{***}$  \\ \\
  \multicolumn{4}{c}{\textit{Utility of friendships}} \\
          & Parameter                                & No SNet Data    & Fixed Net       & Model I          & Model II         & Model III      \\ \hline
    8     & Baseline number of friends               & ---             & ---             & $3.182^{***}$    & $3.144^{***}$    & $2.559^{***}$       \\
    9     & Different sex$^{MP\%}$                   & ---             & ---             & $-0.345^{***}$   & $-0.376^{***}$   & $-0.327^{***}$      \\
    10     & Different grades$^{MP\%}$               & ---             & ---             & $-0.727^{***}$   & $-0.735^{***}$   & $-0.718^{***}$      \\
    11    & Different race$^{MP\%}$                  & ---             & ---             & $-0.394^{***}$   & $-0.277^{***}$   & $-0.337^{***}$      \\
    12    & Reciprocity$^{MP}$                       & ---             & ---             & $0.387^{***}$    & $0.381^{***}$    & $0.354^{***}$       \\
    13    & Degree squared$^{MP}$                    & ---             & ---             & ---              & $0.009^{***}$    & $0.026^{***}$       \\
    14    & Triangles$^{MP\%}$                       & ---             & ---             & $0.019$          & ---              & $-0.056^{***}$      \\
    15    & $\phi^{MP}$                              & ---             & $0.025^{***}$   & $0.028^{***}$    & $0.025^{***}$    & $0.025^{***}$       \\ \hline
    \end{tabular}%
\end{center}
\fignotetitle{Note:} \fignotetext{MP stands for the estimated marginal probability in percentage points and MP$\%$ for estimated marginal probability in percent, relative to the baseline probability. The posterior sample contains $10^5$ simulations before discarding the first $20\%$. Posterior mean outside of the shortest $90/95/99\%$ credible sets is  indicated by $^{*}$/$^{**}$/$^{***}$ respectively.}
\end{table}%

The estimates suggest a substantial role for friends and the home environment in adolescents' decisions to smoke. In particular, one additional friend who is a smoker increases the conditional probability of smoking by $2.5$ ppt.\footnote{Because both friendships and smoking are choices in the model, this parameter should be interpreted with caution. In particular, the estimate cannot be interpreted as the effect on the likelihood of smoking from a randomly assigned friend who is a smoker. In such a case, the individual who is subject to this random assignment may simply drop this friendship as opposed to start smoking.} If $30\%$ of the students in a school smoke, all other things being equal, then an individual is $7.4$ ppt more likely to smoke (row $7$). Also the presence of a smoker in the household increases the likelihood of smoking by $11.5$ ppt. Note that these marginal effects are first order approximations which do not take into account the equilibrium effect after individuals adjust their social norms and friendships. The section with counterfactual experiments presents an equilibrium empirical analysis of the determinants of adolescents' smoking decisions.

Recall that the most central (and challenging) feature of peer influences (in endogenous networks) is the possibilities of externalities and note how the presence of  externalities in friendship formation reveal (row $12$ and $13$) that there is rivalry (exclusivity) for friendship in that the true coefficient on completing a triangle is negative. Furthermore, observe that this is only possible to see once we make the incremental payoff of an extra friendship concave (term degree squared). For if not, the rivalry for friendships motive gets absorbed in the returns to scale motive and the posterior mean of the coefficient from the former is indistinguishable from zero.

A less subtle manifestation of these externalities is the change in the estimate of the price coefficient between the specifications where social network data is not available to the researcher or is kept fixed (the first two columns) and the full model (column $4$). In these cases, the price effect on smoking is incorrectly understated (in fact insignificant). Note that a standard argument signing the omitted variable bias cannot be immediately invoked here. Moreover, in the data the correlation between the (omitted) peer norm and price is of ambiguous sign.\footnote{In more than half of the schools this correlation is positive.} Overall, these findings point to the presence of intricate (non-linear) relationships between smoking, peer norms and exogenous determinants of smoking such as tobacco prices.

\subsection{Model fit}

Using the parameter estimates, a Markov chain of size $10^5$ from the $k$-player dynamic is simulated from which, to reduce the auto dependence, every $1,000$ element is sampled. Table \ref{table:fit} displays selected statistics from the data and from the simulated sample which by design is representative for the stationary distribution. In addition to statistics that are directly targeted by the model's parameters (overall prevalence, density, and reciprocity), statistics which are only indirectly governed by model parameters are reported in tables \ref{table:fit} and \ref{table:fit_mixing}, e.g. maximum degree, certain friendship configurations, mixing etc. Overall the model fits well the smoking decisions and the network features of the data. The only caveat is the proportion of smoking friends as a fraction of the overall friends of a smoker which in the data are $58\%$ while in the model are $46\%$. The most likely reason for this discrepancy is the restriction of $\phi_1=\phi_2$ in the terms $\phi_1 a_ig_{ij}g_{ji}a_j/n$ and $\phi_2 (1-a_i)g_{ij}g_{ji}(1-a_j)/n$. While the proposed approach can handle a more flexible case, the limited data availability precludes the estimation of a more general specification allowing $\phi_1=\phi_2$. 

\begin{table}
\caption{Model fit}
\label{table:fit}
\centering
\begin{tabular}{lcc}
Moment & Model & Data \\ \hline \hline
Prevalence  & 0.325 &  0.308  \\
Density     & 0.017 &  0.017   \\
Reciprocity & 0.345 &  0.328   \\
Avg degree  & 2.155 &  2.199   \\
Max degree out & 7.926 & 8.001 \\
Max degree in  & 8.826 & 9.917 \\
Max reciprocal degree & 4.372 & 4.500 \\
$a_ig_{ij}g_{ji}a_j/n$   & 0.064 & 0.080 \\
$(1-a_i)g_{ij}g_{ji}(1-a_j)/n$ & 0.215 & 0.204 \\
Triangles/n                    & 0.281 & 0.358 \\ \hline
\end{tabular}
\end{table}

\begin{table}[t]
\caption{Fit mixing matrix (model left, data right)}
\label{table:fit_mixing}
\centering
\begin{footnotesize}
\begin{tabular}{llcccc}
  &  & \multicolumn{2}{c}{\axislabel{Nominee}}                                 & \multicolumn{2}{c}{\axislabel{Nominee}} \\
  \multirow{5}{*}{\rotatebox[origin=c]{90}{\axislabel{Nominator}}}
  &            & Smoker               & Nonsmoker                      & Smoker               & Nonsmoker           \\\cmidrule{2-6}
  & Smoker     & \textbf{46\% (53.6)}  &  54\% (62.2)            & \textbf{58\% (69.1)}  &  42\% (48.5)          \\
  & Nonsmoker  & 30\% (63)             & \textbf{70\% (149.9)}   & 29\% (61.5)           & \textbf{71\% (152.4)}\\  \cline{2-6}
\end{tabular}%
\end{footnotesize}
\end{table}

\section{Policy experiments}
\subsection{A. Changes in the price of tobacco}

The estimated model can serve as a numerical prototype to quantify the equilibrium response of various (policy) changes of the decision environment and, more generally, to analyze the determinants of teen smoking.\footnote{To assess the effect of each proposed policy, I consider $10^5$ draws from the stationary distribution of the model.} Table \ref{table:ctrf-price} presents the effects of increases in tobacco prices ranging from $15$ to $150$ cents (in the sample tobacco prices average at $\$1.67$ for a pack). The second through the fifth columns report predictions for the change in overall smoking prevalence in percentage points for various scenarios. I consider the full model, the model when the adjustment of the friendship networks is suppressed, the model where the peer effects coefficients ($\phi$ and $h$) have been set to zero and, finally, the predictions from a model which has been estimated without peer effects altogether.\footnote{Note that the scenario ``PE off''  entails stronger restrictions on individuals' social environments than just keeping friendships fixed. In particular, when simulating individuals' smoking decisions, I keep constant their friendship choices, their friends' smoking statuses, and the average smoking behavior of the population overall (i.e., the number of smokers in the population).}

\begin{table}
\caption{Changes in the price of tobacco}
\label{table:ctrf-price}
\begin{center}
\begin{tabular}{ccccc}
Price increase & Full model & Soc net off & PE off & Model w/o PE \\ \hline \hline
15  &  2.7 &  2.5 & 0.7  & 2.8 \\
30  &  5.1 &  4.5 & 1.3  & 5.4 \\
45  &  7.0 &  6.5 & 2.0  & 7.6 \\
60  &  8.9 &  8.3 & 2.6  & 10.0 \\
75  & 10.4 &  9.7 & 3.1  & 12.0 \\
90  & 11.7 & 11.4 & 4.0  & 13.9 \\
105 & 13.1 & 12.5 & 4.4  & 15.8 \\
120 & 14.1 & 13.5 & 5.0  & 17.6 \\
135 & 15.1 & 14.6 & 5.7  & 19.4 \\
150 & 16.2 & 15.6 & 6.0  & 20.1 \\ \hline
\end{tabular}
\end{center}
\fignotetitle{Note:} The first column shows proposed increases in tobacco prices in cents. The average price of a pack of cigarettes is \$1.67 so that 15 cents is approximately 10\%. The second through fourth columns show the predicted increase in the overall smoking (baseline 32\%) in ppt from the full model, from the model when the friendship network is fixed, and from the model when the coefficients $\phi$ and $h$ are set to zero respectively. The last column shows the predicted change in the overall prevalence produced from a model with no peer effects altogether, i.e. models which is estimated with no peer effects.
\end{table}

\comments{
  15.0000    0.2979    0.0270    0.0251    0.0020    0.0265    0.0066
   30.0000    0.2742    0.0508    0.0458    0.0050    0.0538    0.0133
   45.0000    0.2557    0.0693    0.0655    0.0038    0.0761    0.0199
   60.0000    0.2364    0.0886    0.0825    0.0061    0.0991    0.0255
   75.0000    0.2211    0.1039    0.0970    0.0068    0.1194    0.0312
   90.0000    0.2083    0.1167    0.1141    0.0026    0.1389    0.0384
  105.0000    0.1943    0.1306    0.1254    0.0052    0.1578    0.0438
  120.0000    0.1838    0.1412    0.1351    0.0061    0.1757    0.0495
  135.0000    0.1736    0.1513    0.1457    0.0057    0.1937    0.0572
  150.0000    0.1631    0.1619    0.1565    0.0053    0.2060    0.0598

15 & 7.5 & 4.7 & 0.8 \\
30 & 11.5 & 8.1 & 2.0 \\
45 & 14.2 & 10.7 & 2.7 \\
60 & 16.4 & 12.2 & 3.4 \\
75 & 18.0 & 14.4 & 4.3 \\
90 & 19.6 & 15.7 & 4.9 \\
105 & 20.6 & 16.8 & 5.7 \\
120 & 21.6 & 18.2 & 6.3 \\
135 & 22.6 & 19.3 & 7.0 \\
150 & 23.5 & 20.0 & 7.5 \\ \hline
  
 }

Table \ref{table:ctrf-price} shows that adolescents are sensitive to tobacco prices and that the social interactions, including the response of the social network, amplifies the intended effect of increasing tobacco prices. The baseline smoking in the sample is $32\%$ so that an increase of $\$1.50$ (about $90\%$) reduces the smoking to $16\%$. 
Comparison between model's predictions with and without friendship adjustments (columns two and three) reveals that the equilibrium response of the friendship network to the proposed policies amplifies the intended effect of a price increases by $10\%$, i.e. when high school students are free to adjust their peers, more students stop smoking as a result of a price increase. As we discussed earlier, benefits of the direct effect of increasing tobacco prices when students are free to reconstruct their friendships dominate the benefits of the spillover effects when students are stuck with their friends.\footnote{Figuratively, a price change has two effects on the decision to smoke: the direct effect operates thorough changing individuals' exogenous decision environment and the indirect/spillover effect operates through changing the peer norm which then puts additional pressure on the individuals' to follow the change} In addition, comparison with predictions from the model when peer effects are switched off (columns two and four) suggest that social interactions account for more than $70\%$ of the decrease in smoking following a price increase. Finally, note that if one is to completely discard our equilibrium approach and rely on coarse correlations only,\footnote{This could be implemented with the estimates from column one of table \ref{tab:estimates}, which are obtained from a specification restricting the peer effects coefficients at the stage of estimation.} the predicted smoking decrease is larger. This should not be surprising since the price coefficient in this case absorbs the effect of the (omitted) peer influence (both aggregate and local) and, as we discussed earlier, the bias is negative leading to overstating the magnitude of the price coefficient and whence the predicted effect on smoking.

\comments{
old table with all schools
\begin{tabular}{lccccc} \hline \hline
    Price increase & Model & Fixed friendships & No local PE & No global PE & No PE \\ \hline
    20 cents  & -2.6  & -2.2  & -1.6  & -1.5  & -1.2 \\
    40 cents  & -4.9  & -4.2  & -3.2  & -2.7  & -2.2 \\
    60 cents  & -7.1  & -6.2  & -4.5  & -3.8  & -3.2 \\
    80 cents  & -8.5  & -7.7  & -5.7  & -4.9  & -4.1 \\ \hline
\end{tabular}%
}

\subsection{B. Changes in the racial composition of schools}

Suppose that in a given neighborhood there are two racially segregated schools: ``White School'' consisting of only white students and ``Black School'' consisting of only black students. One would expect that the smoking prevalence in White school is much higher compared to Black school because, in the sample, black high students smoke thee times less than white high school students. Consider a policy aiming to promote racial desegregation, which prevents schools from enrolling more than $x$ percent of students of the same race. If such policy is in place, will students from different races form friendships and will these friendships systematically impact the overall smoking in one or another direction?

To simulate the effect of the proposed policy I consider one of the racially balanced schools in my sample.\footnote{The school has 150 students of which $40\%$ are Whites and $42\%$ are Blacks. It incorporates students from grades 7 to 12.} The Whites and the Blacks from this school serve as prototypes for the White School and Black School respectively.%
\footnote{As an alternative to splitting one school into two racially segregated schools, one could consider two schools from the data that are already racially segregated. However, the only school with a high ratio of Blacks in the sample incorporates students from grades 7 and 8. If this school serves as a prototype for the Black School, then I am faced with two options for the choice of the White School. If the White School incorporates only 7th and 8th graders, then smoking prevalence will be low regardless, since these grades are mostly nonsmoking. Alternatively, if the White School incorporates higher grades, the simulation results will be driven in part by the asymmetry in the population (7th and 8th graders do not make friends with students from higher grades). Consequently, the school that incorporates black students in grades 7 and 8 only cannot properly serve as a prototype for the Black School.} To implement the proposed policy I randomly select a set of students from the White School and a set of students from the Black School and swap them. For example to simulate the effect of a $70\%$ cap on the same-race students in a school, I need to simulate a swap of $30\%$. 

\comments{
\begin{table}[t]
\begin{center}
\caption{Predicted Smoking Prevalence following Same-race Students Cap}
\label{tab:race}%
  \begin{tabular}{cccc} \hline \hline
  $\%$ swapped & White School & Black School & Overall \\ \hline
  0    & 0.322 & 0.035 & 0.173 \\
  10   & 0.295 & 0.041 & 0.163 \\
  20   & 0.281 & 0.061 & 0.167 \\
  30   & 0.224 & 0.069 & 0.144 \\
  40   & 0.223 & 0.095 & 0.157 \\
  50   & 0.175 & 0.162 & 0.168 \\ \hline
\end{tabular}%
\end{center}
\fignotetitle{Note:} \fignotetext{A cap of $x\%$ same-race students is implemented with a swap of $(100-x)\%$ students.}
\end{table}%
}
\begin{table}[t!]
\caption{Predicted Smoking Prevalence following Same-race Students Cap}
\label{table:ctrf-schoolcomposition}
\begin{center}
\begin{tabular}{cccc}
Same-race cap ($\%$) & School White & School Black & Overall \\ \hline \hline
0  & 27.0 & 4.2  & 15.6 \\
10 & 24.5 & 5.7  & 15.1 \\
20 & 20.7 & 8.1  & 14.4 \\
30 & 19.8 & 9.0  & 14.4 \\
40 & 15.9 & 12.0 & 14.0 \\
50 & 14.1 & 13.8 & 13.9 \\ \hline
\end{tabular}
\end{center}
\fignotetitle{Note:} \fignotetext{A cap of $x\%$ same-race students is implemented with a swap of $(100-x)\%$ students.}
\end{table}

\comments{

0  & 27.1 & 6.1  & 16.6 \\
10 & 25.5 & 7.6  & 16.5 \\
20 & 21.3 & 9.9  & 15.6 \\
30 & 21.1 & 10.4 & 15.7 \\
40 & 16.8 & 13.9 & 15.3 \\
50 & 15.5 & 15.4 & 15.5 \\ \hline

     0        0.2829    0.0473    0.1651
   10.0000    0.2606    0.0622    0.1614
   20.0000    0.2169    0.0890    0.1530
   30.0000    0.2087    0.0962    0.1524
   40.0000    0.1679    0.1266    0.1472
   50.0000    0.1542    0.1443    0.1493
   }

Table \ref{table:ctrf-schoolcomposition} presents the simulation results, which suggest that racial composition affects the overall smoking prevalence. 
The first column shows the size of the set of students which is being swapped. The second, third, and forth columns show the simulated smoking prevalences in the White School, Black School, and both, respectively. The table suggest that the overall smoking prevalence in racially segregated schools (the first three rows of the last column) is higher than that of racially desegregated schools (the bottom three rows of the last column). Such a finding provides empirical support for policies promoting racial integration in the context of fighting high smoking rates.

\subsection{C. Cascade effects of an anti-smoking campaign}
The smoking prevalence in the school with the highest smoking rate is $44.7\%$. For this school, I consider the effects of an anti-smoking campaign that can prevent with certainty a given number of students from smoking, e.g. a group of students are invited to a weekend-long information camp on the health consequences of smoking. The camp is very effective in terms of preventing students from smoking; however, it is too costly to engage all students. The question is once the ``treated students'' come back what will happen: will their smoking friends follow their example and stop smoking, or will their friends un-friend them and continue smoking?

\begin{table}
\caption{Spillovers}
\label{table:ctrf-spillovers}
\begin{center}
\begin{tabular}{cccccc}
\multirow{2}{*}{Campaign (\%)}  & \multirow{2}{*}{Smoking}  & \multicolumn{2}{c}{Predicted effect} & Actual & \multirow{2}{*}{Multiplier} \\
& & proportional & fixed network & effect \\
 \hline \hline
-    & 46.4 & - & - & - & - \\
1.0  & 44.7 & 0.5 & 2.9 & 1.7 & 3.7 \\
2.5  & 40.7 & 1.2 & 8.2 & 5.7 & 4.9 \\
5.0  & 35.7 & 2.3 & 12.4 & 10.6 & 4.6 \\
10.0 & 29.0 & 4.6 & 18.1 & 17.4 & 3.7 \\
20.0 & 20.6 & 9.2 & 25.7 & 25.7 & 2.8 \\
30.0 & 16.4 & 13.8 & 29.7 & 29.9 & 2.2 \\
50.0 & 9.7  & 23.0 & 36.3 & 36.7 & 1.6 \\ \hline
\end{tabular}
\end{center}
\fignotetitle{Note:} The first column lists the alternative attendance rates. The second and third columns display the smoking rate and the change in smoking rate respectively. The fourth column computes what would have been the change of the smoking rate if the decrease would be proportional to the intervention, i.e. computes a baseline without peer effects. The last column computes the ratio between the percentage change in the number of smokers and the attendance rate. Note that that attendance is random with respect to the smoking status of the students. If the campaign is able to target only students who are currently smokers, the spillover effects will be even larger.
\end{table}

Table \ref{table:ctrf-spillovers} presents the simulation results, which suggest that an anti-smoking campaign may have a large impact on the overall prevalence of smoking, without necessarily being able to directly engage a large part of the student population.\footnote{The policy is simulated $10^3$ times, where each time a new random draw of attendees is being considered.} In particular, the multiplier factor--the ratio between the actual effect and effect constrained to the treated sub-population--indicated a substantial spillover effects, operating through the social network, from those who attended the camp to the rest of the school.

\comments{
\subsection{E. Improving of home environment}
Finally, table \ref{table:ctrf-homesmokes} presents a quantitative evaluation of, what I consider, the dominant factor for adolescent smoking--family environment.\footnote{The same sample of 9-12 grade students from table \ref{table:ctrf-price} is used.} In $37.7\%$ of the adolescent families, there is a smoker at home. The model predicts that if the family members who smoke at home quit, adolescent smoking will drop by $22.8$ ppt. The magnitude of this estimate suggests that the root of high smoking rates may indeed be adolescents' family environment. Again, disregarding the response of the friendship network to this change will lead to an under estimate of $3$ ppt. Furthermore, comparing these estimates to the direct estimate of the marginal effect of home environment from table \ref{tab:estimates} reveals that peer effects account for almost $50\%$ of the effect in this particular case.

\begin{table}
\caption{Counterfactual experiment improving home environment}
\label{table:ctrf-homesmokes}
\centering
\begin{tabular}{cccc}
Baseline prevalence & Model's prediction & Prediction fixed network & Estimate ME\\ \hline \hline
34.1 & 22.8 & 19.7 & 12.5 \\ \hline
\end{tabular}
\end{table}

}

\section{Concluding remarks}
The premise of this paper is that social norms and behaviors are shaped jointly by a complex process capable of generating a wide range of outcomes. I study this process through the lenses of a novel family of equilibria, $k$-player Nash stability, and I approximate and empirically analyze agents' equilibrium play with the $k$-player dynamic. The empirical application of the proposed framework to adolescents' friendship selection and decisions to smoke makes the first step in understanding the mechanisms and the opportunities for public policy arising from the non-linear relationship between the social norms and individuals' behaviors theorized by \cite{graham2014complementarity} and experimentally discovered in \cite{carrell2013natural}. Overall this research formulates an avenue to study the complementarities and coordination in social networks with accent on the possibility for multiple equilibrium outcomes.

\newpage
\appendix
\section{Proofs}

\begin{proof}[Lemma \ref{lemma:equilibrium_equivalence} (on p. \pageref{lemma:equilibrium_equivalence})] \label{proof_lemma:equilibirum_equivalence}
For part $(i)$, if $S$ is $k$-PS then $S|_{I_r}$ is a Nash equilibrium of $\Gamma |_{I_r}$ for any $I_r$ such that $|I_r| = k$. If no player has incentive to deviate in $\Gamma |_{I_r}$ then no player have incentive to deviate in $\Gamma |_{I_r'}$ where $I_r'\subseteq I_r$ and $r'<r$. This establishes that any partition with maximal component of size $k$ is Nash stable. The converse follows mutatis mutandis. Part $(ii)$ follows from what we just established. If a state is partition Nash stable with respect to any partition with maximum component of size $k$, it will certainly be partition Nash stable with respect to a finer partition with smaller maximum component.
\qed
\end{proof}

\begin{proof}[Proposition \ref{prop:existence} (on p. \pageref{prop:existence})] \label{proof:prop:existence}
The argument relies on showing that the normal form game is a game with potential (\citealp{monderer_shapley_96}). This property has implication not only for the existence results but also for the adaptive dynamic analysis.

\begin{claim} \label{prop:potential}
The preferences given in (\ref{util1}) and (\ref{util2}) are summarized with the {\normalfont potential function} $\Pfn : \Gn \times \mathbf{X}_n \longrightarrow \mathbb{R}$ :
\begin{eqnarray} \label{pfn}
\Pfn(S,X)   &=&  \sum_{i} a_{i} v(X_{i})
  +\frac{1}{2} h \sum_{i}\sum_{j}a_{i}a_{j} 
  +\frac{1}{2}\phi \sum_{i}\sum_{j}a_{i}a_{j}g_{ij}g_{ji}
   \\ \notag
  & & + \sum\limits_{i}\sum\limits_{j}g_{ij}w\left(X_{i},X_{j}\right)
  +\frac{1}{2}m\sum\limits_{i}\sum\limits_{j}g_{ij}g_{ji} \\ \notag
  & & + q\sum\limits_{i}\sum\limits_{j}\sum\limits_{k}\left(\frac{g_{ij}g_{jk}g_{ki}}{3}\right)
\end{eqnarray}
\end{claim}

The claim is equivalent to following conditions:.

\bit{Condition A.} For any $i$, $S_{-i}$ and $X$
\begin{equation} \label{condition_A} 
  \Pfn(S',X)-\Pfn(S,X) = u_i(S',X) - u_i(S,X)
\end{equation}
where $S,S' \in \Gn$ are defined as $S=(a_i=0,S_{-i})$ and $S'=(a_i=1,S_{-i})$.

\bit{Condition B.} For any $i\neq j$, $S_{-ij}$ and $X$
\begin{equation} \label{condition_B} 
  \Pfn(S',X)-\Pfn(S,X) = u_i(S',X) - u_i(S,X)
\end{equation}
where $S,S' \in \Gn$ are defined as $S=(g_{ij}=0,S_{-ij})$ and $S'=(g_{ij}=1,S_{-ij})$.

The verification of these conditions is a matter of trivial algebra. \qed
\end{proof}

\begin{proof}[Proposition \ref{prop:dynamics} (on p. \pageref{prop:dynamics}] \label{proof:prop:dynamics}
That any k-player Nash stable state is absorbing follows from the definition of $\KNS$. The second part follows from the following claim:
\begin{claim}
 The sequence of the values of the potential $\Pfn_t$ induced by the outcome of a meeting process of dimension k is a submartingale, i.e. $E[\Pfn_{t+1}|S_t ] \geq \Pfn_t$.
\end{claim}
So that $\{\Pfn_t\}$ converges almost surely. Since the network is of finite size it follows that $\{\Pfn_t\}$ is constant for large $t$. Because of assumption \ref{asm:meet1} this can happen only if $S_t \in \KCS$.
\qed
\end{proof}

\begin{proof}[Theorem \ref{thm:stationarydistribution} (p. \pageref{thm:stationarydistribution})] \label{proof:thm:stationarydistribution}
Note that the $k$-PD induces a finite state Markov chain which is well-behaved: (i) irreducible, (ii) positive recurrent, and (iii) aperiodic. Then, the first part of the theorem follows from standard results on convergence of Markov chains. To describe the $k$-PD we need to integrate out the meeting process and obtain the one step transition probabilities $\Pr(S'|S)$ for $S,S'\in \SSn$. In the end, we will show that $\Pr(S'|S)\exp\{\Pfn(S)\} = \Pr(S|S')\exp\{\Pfn(S')\}$ so that, when properly normalized, $\exp\{\Pfn(S)\}$ is the stationary distribution.

Let $\dSSn$ be the set of all probability distributions on $\SSn$. Because $\SSn$ has no natural ordering, one can think of $p \in \dSSn$ as a function $p : \SSn \rightarrow \R$ such that $p(S)\geq 0$ and $\sum p(S) = 1$.\footnote{The set $\dSSn$ can be extended and suitably equipped with an inner product, and shown to be a Hilbert space, e.g. for a measure $\pi$ and $f,g:\SSn \rightarrow \R$, let $(f,g)=\int fg d\pi$ and $\|f\|=\sqrt{(f,f)}$. Then, it can be shown that $L^2(\pi)=\{f:\|f\|<\infty\}$ is a linear space, $(.,.)$ satisfy the axioms of inner product, and $L^2(\pi)$ is topologically closed.} Let $T_k$ be the one step transition probability associated with $k$-PD. $T_k$ a linear operator on $\dSSn$.

Consider the case when $\Pr(S'|S)>0$.\footnote{Out assumptions guarantee that $\Pr(S',S)>0$ iff $\Pr(S,S')>0$. Thus, if $\Pr(S'|S)=0$ then $\Pr(S|S')=0$ and, trivially, $\Pr(S'|S)\exp\{\Pfn(S)\} = \Pr(S|S')\exp\{\Pfn(S')\}$ holds.} For fixed $S,S'\in \SSn$ let $\MK_{S'|S}$ be the set of all possible meeting outcomes which can result in state transitioning from $S$ to $S'$. For example, if $S$ and $S'$ differ in the status of $a_{ii}$ then $\MK_{S'|S}=\{(a_{ii},I_{k-1})| I_{k-1}=\{i_1,i_2\ldots\i_{k-1}\}, i_j\neq i \}$. 
Recall that ${\NS}_k(S,\mu)\subset \SSn$ denotes the set of all possible outcomes of the meeting $\mu$ following a state $S$. The proof follows from the following observations:\footnote{The formal proof involves basic reasoning revolving around the intuition behind these observations. The challenging part is to state and interpret the lemma. Below I provide the interpretation while omitting the proof which is available upon request.}

\begin{lemma}
\label{neibourhoods}
Let $S,S'\in \SSn$ and $\mu=(i,I_{k-1})$. Then
\begin{enumerate} [topsep=0ex,parsep=0ex,itemsep=0ex,leftmargin=0.7cm,label=(\roman*)]
\item $\MK_{S'|S}=\MK_{S|S'}$ for all $S,S'\in \SSn$.
\item $S' \in \NS_k(\mu,S)$ iff $S \in \NS_k(\mu,S')$.
\item If $S' \in \NS_k(\mu,S)$ then $\NS_k(\mu,S)=\NS_k(\mu,S')$.
\end{enumerate}
\end{lemma}

Part $(i)$ asserts that each meeting which can result in the state moving from $S$ to $S'$ may result in, provided the starting state were $S'$, the state moving from $S'$ to $S$. Part $(ii)$ states that if there is a meeting which can result in $S$ transiting to $S'$, then there is a meeting which can result in $S'$ transiting to $S$ (note that by $(i)$ we know this is the same meeting). Finally, part $(iii)$ notes that if a meeting $\mu$ could result in $S$ transiting to $S'$, not only $\mu$ can result in $S'$ transiting to $S$ (by $(i)$), but also the set of all feasible states following $\mu$ and $S$ coincides with the set of all feasible states following $\mu$ and $S'$.
From lemma \ref{neibourhoods}, the one step transition probability can be written as:
\begin{eqnarray}
\Pfn(S) \Pr(S'|S) & = &\Pfn(S) \sum_{\mu \in \MK_{S'|S}} \Pr(\mu) \frac{\exp\{u_{i}(S')\}}{\sum_{\hat{S} \in \NS_k(\mu,S)}\exp\{u_{i}(\hat{S})\}} \\
& = &\Pfn(S)\sum_{\mu \in \MK_{S|S'}} \Pr(\mu) \frac{\exp\{\Pfn(S')\}}{\sum_{\hat{S} \in \NS_k(\mu,S)}\exp\{\Pfn(\hat{S})\}}\\
& = &\Pfn(S')\sum_{\mu \in \MK_{S|S'}} \Pr(\mu) \frac{\exp\{\Pfn(S)\}}{\sum_{\hat{S} \in \NS_k(\mu,S ')}\exp\{\Pfn(\hat{S})\}}\\
& = & \Pfn(S) \Pr(S,S')
\end{eqnarray}

Formally this completes the proof and, for an illustration, I write down the transition probability for the special case when $S$ and $S'$ agree on all $\{g_{ij}\}_{i\neq j}$ but differ in $a_{ii}$ for some $i$, say $S=(a_{ii}=0,S_{-i})$ and $S'=(a'_{ii}=1,S_{-i})$ (here $S_{-i}=\{g_{ij}\}_{i\neq j}$). Then we have 
\begin{equation*}
\Pr(S'|S)=\sum_{\mu \in \MK_{|ii}} \Pr(\mu) \frac{\exp\{u_{i}(S')\}}{\sum_{\hat{S} \in \NS_k(\mu,S)}\exp\{u_{i}(\hat{S})\}}
\end{equation*}
where $\MK_{|ii}$ is the set of all possible meeting tuples $(i,I_{k-1})$ where player $i$ meets different $I_{k-1}\subset \{1,\ldots, i-1,i+1,\ldots,n\}$. Note that $|\MK_{|ii}|={n-1 \choose k-1}$. In the case where all meeting are equally likely and individuals are indifferent to all outcomes (i.e. $u_i$ is a constant), the above reduces to
\begin{equation*}
\Pr(S'|S)={n-1 \choose k-1} \frac{1}{n{n-1 \choose k-1}} \frac{1}{2^k}=\frac{1}{n 2^k}
\end{equation*}

\qed
\end{proof}

\begin{proof}[Theorem \ref{thm:convergencespeed} (p. \pageref{thm:convergencespeed})]
\label{proof:thm:convergencespeed}
Because there is no natural ordering of $\SSn$, I will use functions as opposed to vectors in the eigen problem. For $I\subset \{1,\ldots,n\}\times\{1,\ldots,n\}$, define $\e_I : \SSn \rightarrow \R$ as
\begin{equation}
\e_I(S) = \prod_{i\neq j\in I} (-1)^{g_{ij}} \prod_{i=j\in I} (-1)^{a_{ij}}
\end{equation}
with $\e_\emptyset(S) =1$ for all $S$. Next, define 
\begin{equation}
\lambda_{k,I} = \frac{\sum_{i\in\{(i,i)\notin I\}}{n-1-|I_i| \choose k-1}}{n{n-1 \choose k-1}}
\end{equation}
where $I_i = \{j:(i,j)\in I, i\neq j\}$

\begin{lemma} There are $2^{n^2}$ pairs of $(\lambda_{k,I},\e_{k,I})$ such that
\begin{enumerate} [topsep=0ex,parsep=0ex,itemsep=0ex,leftmargin=0.7cm,label=(\roman*)]
\item $\sum_S \e_{k,I}(S) \e_{k,I'}(S)=0$ if $I\neq I'$ and $\sum_S \e_{k,I}(S) \e_{k,I}(S)=2^{n^2}$
\item For any $S\in \SSn$
\begin{eqnarray}
\sum_{S'} \Pr(S'|S)\e_I(S')= \frac{\sum_{(i,i)\notin I}{n-1-|I_i| \choose k-1}}{n{n-1 \choose k-1}} \e_I(S).
\end{eqnarray}
\end{enumerate}
\end{lemma}

\noindent
More concisely $T_k \e_I=\lambda_I \e_I$.
 
The first part of the lemma is trivial to verify. For the second part, we can write:
\begin{eqnarray}
\sum \Pr(S'|S)\e_I(S') & = & \sum_{S'} \sum_\mu \Pr(\mu) \Pr(S'|S,\mu) \e_I(S') \\
& = & \sum_{S'} \sum_{\mu\in\{\mu\cap I=\emptyset\}} \Pr(\mu) \Pr(S'|S,\mu) \e_I(S') +  \\
& & + \sum_{S'} \sum_{\mu\in\{\mu\cap I \neq \emptyset\}} \Pr(\mu) \Pr(S'|S,\mu) \e_I(S') \\ \label{eq:transition}
& =&  \sum_{S'} \sum_{\mu\in\{\mu\cap I=\emptyset\}} \Pr(\mu) \Pr(S'|S,\mu) \e_I(S')
\end{eqnarray}
because 
$
\sum_{\mu\in\{\mu\cap I \neq \emptyset\}} 
\sum_{S'\in \NS_k(S,\mu)}
\Pr(S'|S,\mu) \e_I(S') = E[\e_I(S')|S,\mu\cap I\neq \emptyset] =0
$. This summation involves $2^k$ terms of $\NS_k(S,\mu)$. It is easy to see that for half of these terms $\e_I(S')=\e_I(S)$ and for the other half $\e_I(S')=-\e_I(S)$.

Finally, note that $\e_I(S)=\e_I(S')$ provided $\mu\in\{\mu\cap I=\emptyset\}$ so that for (\ref{eq:transition}) we can write
\begin{eqnarray}
\sum \Pr(S'|S)\e_I(S')
& = & \e_I(S) \Pr(\mu) \sum_{S'} \sum_{\mu\in\{\mu\cap I=\emptyset\}} \Pr(S'|S,\mu) \\
& = & \e_I(S) \frac{1}{n{n-1 \choose k-1}} \sum_{i\in\{(i,i)\notin I\}} {n-1-|I_i| \choose k-1}
\end{eqnarray}
because, by assumption, $\Pr(\mu) = \frac{1}{n{n-1 \choose k-1}}$ and $\Pr(S'|S,\mu)= \Pr(S''|S,\mu)$ for $S',S''\in \NS_k(S,\mu)$. With this the proof of the lemma is complete. To complete the proof of the theorem note that $\lambda_{k,I}$ are decreasing in $|I|$, so that the (second) smallest $\lambda_{k,I}$ is achieved when $I=\{(i,j)\}$ with $i\neq j$.
\qed
\end{proof}

\begin{proof}[Theorem  \ref{thm:equlibriumranking} (p. \pageref{thm:equlibriumranking}] \label{proof:thm:equilibriumranking}
The proof follows immediately from the expression for the stationary distribution obtained in theorem \ref{thm:stationarydistribution}.
\qed
\end{proof}

\begin{proof}
[Proposition  \ref{prop:qre} (p. \pageref{prop:qre}]
\label{proof:prop:qre}
I will proceed by a way of contradiction. Consider a network with $n=2$ players and suppose all coefficients except $m$ are set to zero. In addition, consider the subspace $\overline{\Sn}$ of $\Sn$ consisting of the product of the linking strategies only: $$\overline{\Sn}=\{g_{12}=0,g_{12}=1\} \times \{g_{21}=0,g_{21}=1\}.$$ Table \ref{tab:non_factorizability} shows the distribution $\pi$ conditional on $a_1=a_2=0$ on $\overline{\Sn}$ up to a normalizing factor. Clearly this matrix is full rank (its determinant is nonzero provided $w\neq0$) and thus cannot be factored into two independent marginals (i.e., play where individuals 1 and 2 randomize independently).
\begin{table}
\begin{center}
\caption{Example non-factorizability of $\pi$.}
 \label{tab:non_factorizability}
\begin{tabular}{c|c|c|}
\multicolumn{1}{c}{}& \multicolumn{1}{c}{$g_{21}=0$} & \multicolumn{1}{c}{$g_{21}=1$}  \\ \cline{2-3}
$g_{12}=0$ & $1$  &  $1$        \\ \cline{2-3}
$g_{12}=1$ & $1$  &  $\exp\{w\}$ \\ \cline{2-3}
\cline{2-3}
\end{tabular}
\end{center}
\end{table}
\qed
\end{proof}

\begin{proof}
[Proposition  \ref{prop:varying_mh} (p. \pageref{prop:varying_mh}]
\label{proof:prop:varying_mh}

For fixed $S,S'\in \SSn$ let $\KK_{S'|S}\subset\{2,3,\ldots, n\}$ be the set of all possible meeting sizes consistent with transition from $S$ to $S'$ of the $k$-PD. Recall that, for fixed $k$, $\MK_{S'|S}$ is the set of all possible meeting outcomes which can result in state transitioning from $S$ to $S'$. The argument bellow follows from lemma \ref{neibourhoods}, together with the observation that $\KK_{S'|S}=\KK_{S|S'}$. Indeed, the unconditional proposal $Q$ from the algorithm in table \ref{table:algorithm} can be written as:

\begin{eqnarray}
Q(S'|S) & = &\sum_{k\in\KK_{S'|S}} p_k(k)\sum_{\mu \in \MK_{S'|S}} \Pr(\mu) \frac{1}{|\NS_k(\mu,S)|} \\
& = &\sum_{k\in\KK_{S|S'}} p_k(k) \sum_{\mu \in \MK_{S|S'}} \Pr(\mu) \frac{1}{|\NS_k(\mu,S)|}\\
& = &\sum_{k\in\KK_{S|S'}} p_k(k) \sum_{\mu \in \MK_{S|S'}} \Pr(\mu) \frac{1}{|\NS_k(\mu,S ')}\\ \notag
& = & Q(S|S')
\end{eqnarray}
\qed
\end{proof}

\section{Data}
\subsection{Add Health Data}
This research uses data from Add Health, a program project directed by Kathleen Mullan Harris and designed by J. Richard Udry, Peter S. Bearman, and Kathleen Mullan Harris at the University of North Carolina at Chapel Hill, and funded by grant P01-HD31921 from the Eunice Kennedy Shriver National Institute of Child Health and Human Development, with cooperative funding from 23 other federal agencies and foundations. Special acknowledgment is due Ronald R. Rindfuss and Barbara Entwisle for assistance in the original design. Information on how to obtain the Add Health data files is available on the Add Health website (http://www.cpc.unc.edu/addhealth). No direct support was received from grant P01-HD31921 for this analysis.

\subsection{Sample selection and sample statistics}


This research uses data from Wave I of Add Health. The in-home questionnaire contains $44$ sections collecting a wide array of information about adolescents. In particular, the data contain information about adolescents' friendship networks. Each respondent is asked to nominate up to five of her best male and female friends. If individual A nominates individual B as a friend, this does not imply that B nominates A. Indeed, only $36\%$ of the friendships in the saturated sample are reciprocal.\footnote{Also, both the in-school and the in-home questionnaires contain data about best 5 male and best 5 female friendships. For about $26\%$ of the cases in the in-home sample, however, the interviewer asked only about best 1 male and best 1 female friends. In these cases, to prevent the friendship network from being truncated, I use friendship nominations data from the in-school sample whenever available.}

In addition to the friendship network data, I use demographic data for the adolescents (age, gender, grade, and race), for their home environments (presence of smoker in the household, pupil's income and allowances, and mother's education), and data for their smoking behavior. The adolescent's smoking status is deduced from the question, ``During the past $30$ days, on how many days did you smoke cigarettes?'' and if the answer was one or more days, the student's smoking status is set to positive. Because all of the students in the saturated sample were eligible for in-home interview, I have detailed information about student friends as well.

As pointed earlier the schools from the saturated sample (16 schools out of 80) were illegible for exhaustive survey. Since the size of the schools from this sample ranges from $20$ to more than $1500$, the smallest and the largest schools are dropped. After this still the largest school in the sample enrolls more than 3 times more students compared to the second largest. To maintain sample observations of comparable size (each school is an observation), the largest school is split into grades $9$, $10$, $11$, and $12$ and, for this school, each grade is treated as a separate network.\footnote{Less than $20\%$ of the friendships are inter-grade so that this split does not affect substantially the friendship network.} In addition, schools with students only in grades $7$ and $8$ are dropped. Table \ref{table:descriptive_stats} shows selected descriptive statistics for the estimation sample.



\begin{table}[t]
\begin{center}
\caption{Descriptive Statistics for the Final Sample}
  \begin{tabular}{lccc}
    \hline \hline
             & Overall & Min      & Max   \\ \hline
  Students   & 1634  & 44         & 234   \\
  Smoking    & 0.324 & 0.045      & 0.536 \\
  Male       & 0.504 & 0.346      & 0.581 \\
  Whites     & 0.861 & 0          & 0.989 \\
  Blacks     & 0.087 & 0          & 0.975 \\
  As-Hi-Ot   & 0.052 & 0          & 0.373 \\
  Price       & 166.8 & 137.3     & 220.1 \\
  Mom edu     & 0.747 & 0.684     & 0.939 \\
  HH smokes   & 0.452 & 0.092     & 0.609 \\
  Avg friends & 2.432 & 0.273     & 3.503 \\
  \hline
  \end{tabular}%
\label{table:descriptive_stats}
\end{center}
\fignotetitle{Note:} \fignotetext{The final sample contains students from 12 high school networks.}
\end{table}%

\section{Background on tobacco smoking}

Tobacco is the single greatest preventable cause of death in the world today.\footnote{The World Health Organization, \emph{Report on the Global Tobacco Epidemic} ($2008$). The statistics for the U.S. are compiled from reports by the Surgeon General ($2010$), National Center for Health Statistics ($2011$), and Monitoring the Future (2011).} In the United States alone, cigarette smoking causes approximately $443,000$ deaths each year (accounting for one in every five deaths) and imposes an economic burden of more than $\$193$ billion a year in health care costs and loss of productivity. Approximately $1$ million young people under $18$ years of age start smoking each year; about $80\%$ of adults who are smokers started smoking before they were $18$ \citep{kessler_etal_96,liang_chaloupka_nichter_clayton_01}. Despite an overall decline in smoking prevalence from $2005$ to $2010$, when the percentage of current smokers decreased from $20.9\%$ to $19.3\%$, the reduction in teen smoking has been less pronounced. In fact, the proportions of $8$th and $10$th graders who smoke increased slightly in $2010$. As with many human behaviors, social interactions (peer influence) have often been pointed to as a major driving force behind adolescent smoking choices.

\newpage
\nocite{amemiya_81,
        Blume_1993,
        bisin_moro_topa_11,
        card_giuliano_11,
        case_katz_91,
        Cournot_1838,
        currarini_jackson_pin_09,
        degiorgi_pellizzari_redaelli_10,
        gilles_sarangi_04,
        gilleskie_zhang_10,
        heckman_78,
        jackson_08,
        Hops20-Sandia,
        kandori_mailath_rob_93,
        krauth_05, krauth_05b,
        liang_chaloupka_nichter_clayton_01,
        lavy_schlosser_11,
        manski_93,
        moffitt_01,
        monderer_shapley_96,
        rosenthal_73}
\bibliographystyle{aer}
\addcontentsline{toc}{chapter}{Bibliography}
\bibliography{bib01}
\end{document}